\documentclass[acmsmall]{acmart}

\setcopyright{cc}
\setcctype{by}
\acmJournal{PACMMOD}
\acmYear{2026} \acmVolume{4} \acmNumber{3 (SIGMOD)} \acmArticle{161}
\acmMonth{6} \acmDOI{10.1145/3802038}

\usepackage{amsmath,amssymb,amsfonts}
\usepackage{graphicx}
\usepackage{textcomp}
\usepackage{color}
\usepackage{makecell}
\usepackage{multirow}
\usepackage{enumitem}
\usepackage{booktabs}
\usepackage{bbding}
\usepackage{colortbl}
\usepackage{xcolor}
\usepackage{setspace}
\usepackage{pifont}
\usepackage{caption}
\usepackage{subcaption}
\usepackage{url}
\usepackage{listings}
\usepackage[ruled,linesnumbered,vlined]{algorithm2e}
\usepackage{float}
\newcommand{\sqlflex}{SQLFlex}
\newcommand{\sqlglot}{SQLGlot}
\newcommand{\pglast}{\texttt{pglast}}
\newcommand{\sqlcheck}{SQLCheck}
\newcommand{\sqlfluff}{SQLFluff}
\newcommand{\sqlancer}{SQLancer}
\newcommand{\sqlninetytwo}{SQL-92}
\newcommand{\sqless}{SQLess}
\newcommand{\eg}{\emph{e.g.,}}
\newcommand{\ie}{\emph{i.e.,}}

\newcommand{\rewriter}{\emph{Rewriter}}
\newcommand{\printer}{\emph{Pretty-printer}}

\newcommand{\codeIn}[1]{\texttt{\small\detokenize{#1}}}

\definecolor{codeblue}{RGB}{0, 119, 182}
\lstdefinestyle{myStyle}{ 
    commentstyle=\color{codeblue},
    basicstyle=\ttfamily\scriptsize,
    breakatwhitespace=false,
    breaklines=true,
    frame=lines,
    keepspaces=false,                 
    numbers=none,       
    numbersep=5pt,                  
    showspaces=false,                
    showstringspaces=false,
    showtabs=false,                  
    tabsize=2
}
\lstdefinestyle{custom}{
  basicstyle=\ttfamily\footnotesize,
  keywordstyle=\bfseries,
  morekeywords={SELECT,FROM,WHERE,INNER,LEFT,RIGHT,FULL,CROSS,GROUP,BY,HAVING,DISTINCT,JOIN},
  emph={INNER,LEFT,RIGHT,FULL,CROSS,DISTINCT, ALL, AND, OR},
  emphstyle=\bfseries,
literate={<}{{{\textbf{<}}}}1
       {/}{{{\textbf{/}}}}1,
  showstringspaces=false,
  numbers=left,
  numberstyle=\footnotesize,
  stepnumber=1,
  numbersep=10pt,
  tabsize=2,
  captionpos=t,
  frame=none,
  rulecolor=\color{black},
  escapeinside={(*@}{@*)},
  xleftmargin=1.5em,
  belowskip=-1em
}

\lstdefinestyle{tsql}{
  basicstyle=\ttfamily\footnotesize,
  keywordstyle=\bfseries,
  morekeywords={SELECT,FROM,WHERE,INNER,LEFT,RIGHT,FULL,CROSS,GROUP,BY,HAVING,DISTINCT,JOIN, AS, AND, "<"},                               
  emphstyle=\color{purple}\bfseries,                    
  showstringspaces=false,
  numbers=left,
  numberstyle=\footnotesize,
  stepnumber=1,
  numbersep=10pt,
  tabsize=2,
  captionpos=t,
  frame=none,
  rulecolor=\color{black},
  escapeinside={(*@}{@*)},
  moredelim=**[is][\color{red}]{@}{@},
  xleftmargin=1.5em,
  belowskip=-1em,
  aboveskip=0em
}

\lstdefinestyle{inline}{
  basicstyle=\ttfamily\footnotesize,
  keywordstyle=\bfseries,
  morekeywords={SELECT,FROM,WHERE,INNER,LEFT,RIGHT,FULL,CROSS,GROUP,BY,HAVING,DISTINCT,JOIN, CREATE, TABLE, NOT, NULL, AS, AND, "<"},                               
  emphstyle=\color{purple}\bfseries,                    
  showstringspaces=false,
  numbers=none,
  numberstyle=\footnotesize,
  stepnumber=1,
  numbersep=10pt,
  tabsize=2,
  captionpos=t,
  frame=none,
  rulecolor=\color{black},
  escapeinside={(*@}{@*)},
  moredelim=**[is][\color{red}]{@}{@},
  xleftmargin=1.5em,
  belowskip=0em,
  aboveskip=0.5em
}

\definecolor{codebg}{RGB}{245,245,248} 

\lstdefinestyle{python}{
  language=Python,
  basicstyle=\ttfamily\footnotesize,
  keywordstyle=\bfseries\color{black},
  commentstyle=\itshape\color{teal},
  stringstyle=\color{blue},
  showstringspaces=false,
  numbers=left,
  numberstyle=\footnotesize,
  stepnumber=1,
  numbersep=10pt,
  tabsize=4,
  frame=none,
  xleftmargin=1.5em,
  aboveskip=0em,
  belowskip=-0.5em,
  escapeinside={(*@}{@*)},
  emph={find,transform},                               
  emphstyle=\color{orange}\bfseries,                    
}

\lstdefinestyle{python_prompt}{
  language=Python,
  basicstyle=\ttfamily\footnotesize,
  keywordstyle=\bfseries\color{black},
  commentstyle=\itshape\color{teal},
  stringstyle=\color{blue},
  showstringspaces=false,
  numbers=none,
  stepnumber=1,
  numbersep=10pt,
  tabsize=4,
  escapeinside={(*@}{@*)},
  emph={find,transform},                               
  emphstyle=\color{orange}\bfseries,                    
backgroundcolor=\color{codebg},
  rulecolor=\color{black!20},
    frame=single,
    frameround=ttt,
    belowskip=0em,
    aboveskip=0em
}

\lstdefinestyle{boxed}{
  backgroundcolor=\color{codebg},
  basicstyle=\ttfamily\footnotesize,
  frame=single,              
  frameround=tttt,           
  rulecolor=\color{black!20}, 
  breaklines=true,
  breakautoindent=false,
  breakindent=0pt,
  columns=fullflexible,
  belowskip=0em,
  aboveskip=0em
}

\begin{document}
\title{Dialect-Agnostic SQL Parsing via LLM-Based Segmentation}

\author{Junwen An}
\orcid{0009-0003-1768-7443}
\affiliation{
  \institution{National University of Singapore}
  \country{Singapore}
}
\email{junwenan@u.nus.edu}

\author{Kabilan Mahathevan}
\orcid{0009-0003-4980-4886}
\affiliation{
  \institution{Virgina Tech}
  \city{Blacksburg, VA}
  \country{USA}
}
\email{kabilan@vt.edu}

\author{Manuel Rigger}
\orcid{0000-0001-8303-2099}
\affiliation{
  \institution{National University of Singapore}
  \country{Singapore}
}
\email{rigger@nus.edu.sg}


\begin{abstract}

SQL is a widely adopted language for querying data, which has led to the development of various SQL analysis and rewriting tools. However, due to the diversity of SQL dialects, such tools often fail when encountering unrecognized dialect-specific syntax. While Large Language Models (LLMs) have shown promise in understanding SQL queries, their inherent limitations in handling hierarchical structures and hallucination risks limit their direct applicability in parsing. To address these limitations, we propose SQLFlex, a novel query rewriting framework that integrates grammar-based parsing with LLM-based segmentation to parse diverse SQL dialects robustly. Our core idea is to decompose hierarchical parsing to sequential segmentation tasks, which better aligns with the strength of LLMs and improves output reliability through validation checks. Specifically, SQLFlex uses clause-level segmentation and expression-level segmentation as two strategies that decompose elements on different levels of a query. We extensively evaluated SQLFlex on both real-world use cases and in a standalone evaluation. In SQL linting, SQLFlex outperforms SQLFluff in ANSI mode by 63.68\% in F1 score while matching its dialect-specific mode performance. In test-case reduction, SQLFlex outperforms SQLess by up to 10 times in simplification rate. In the standalone evaluation, it parses 91.55\% to 100\% of queries across eight distinct dialects, outperforming all baseline parsers. We believe SQLFlex can serve as a foundation for many query analysis and rewriting use cases.


\end{abstract}

\maketitle

\section{Introduction}

Relational Database Management Systems (RDBMS) are among the most widely adopted data management platforms, with Structured Query Language (SQL) as the main interface for interacting with them. A core application of SQL is querying data. As a result, many important query-related tasks have emerged, which typically involve query analysis and rewriting. For example, SQL linting tools analyze queries for anti-patterns~\cite{dintyalaSQLCheckAutomatedDetection2020, SQLFluff}, and may rewrite queries to fix such issues. Formally, we define \emph{query rewriting} as transforming an input query $Q$ into a new query $Q'$ that satisfies a target objective $O$. \emph{Query analysis} examines queries without modifying them. Since most rewriting begins with analysis, we use \emph{rewriting} to refer to both for simplicity. Under this formulation, a wide range of applications beyond linting involve query rewriting, such as query reduction~\cite{lin2024SQLessDialectAgnosticSQL}, DBMS testing~\cite{jiang24eet, Rigger2020TLP, liuAutomaticDetectionPerformance2022}, and SQL grading~\cite{chandra2015DataGeneration, chandra2019AutomatedGradingSQL}.




A major challenge in query rewriting is the diversity of SQL dialects, which leads to syntactic incompatibilities across systems~\cite{zhou2025cracking, zhong2024UnderstandingReusingTest}. Most query rewriting tools follow a common workflow, but dialect-specific syntax frequently leads to failures. Specifically, these tools first parse an input SQL query with a grammar-based parser into an \emph{Abstract Syntax Tree} (AST). The tools then analyze or modify the AST, and finally generate a query from the AST. The grammar-based parsers used in these tools often fail when encountering dialect-specific syntax, limiting their ability to support a wide range of DBMSs. For example, SQLFluff~\cite{SQLFluff}, a popular SQL linter, has more than 200 dialect-related open issues in its repository.


Grammar-based SQL parsers typically operate on a fixed set of grammar rules, and fail when encountering syntax unrecognized by the grammar. Some tools integrate grammar rules from multiple dialects into a single parser, allowing them to parse queries from multiple dialects into a unified AST. Notable examples include SQLFluff~\cite{SQLFluff} and \sqlglot{}~\cite{SQLGlot}. Although effective for the dialects they support, these tools require substantial manual effort to support new dialects. For example, although \sqlglot{} supports over 30 dialects, its entire parser codebase exceeds 20,000 lines, and adding support for a new dialect often requires over 1,000 additional lines. Additionally, existing dialects may introduce new features over time, requiring continued effort from such parsers' developers to support them. For instance, DuckDB recently integrated the \codeIn{MATCH_RECOGNIZE} row pattern matching feature~\cite{lambrecht2025democratize}.
As a task-specific approach that tackles the SQL dialect problem, \sqless{}~\cite{lin2024SQLessDialectAgnosticSQL}, a test case reducer, proposes an adaptive parser that attempts to generate new grammar rules when encountering dialect-specific features automatically. However, our experiments show that the generated rules fail to produce ASTs that can be interpreted by rewriting rules. Consequently, it remains a challenge to develop an automatic dialect-agnostic query parsing approach that generates an AST representation suitable for a wide range of query analysis and rewriting tasks.

%

Recent advances in Large Language Models (LLMs) have shown promise in SQL-related tasks~\cite{gao2024TextSQLEmpoweredLarge, li2024DawnNaturalLanguage, liCanLLMAlready}. Although LLMs demonstrate SQL understanding abilities, using them to directly generate an AST poses challenges. Their autoregressive nature makes them less effective at handling complex hierarchical structures, such as ASTs~\cite{he2024HierachyEncodersLLM, nandi-etal-2025-sneaking, jiang2025hibenchbenchmarkingllmscapability}. Additionally, end-to-end generation is prone to hallucinations~\cite{li-etal-2023-llm-constituency-parsing}, especially since no universal SQL grammar exists to validate outputs across dialects. We demonstrate these difficulties empirically in our evaluation.

In this paper, we take the first step toward addressing the limitations of grammar-based parsers and the drawbacks of using LLMs naively for dialect-agnostic query parsing. We aim to combine the accuracy of grammar-based parsing with the flexibility of an LLM-based segmenter to build an AST. To this end, we propose \sqlflex{}, a novel query rewriting framework that adopts this hybrid query parsing approach. While grammar-based parsers perform well at handling inputs that conform to predefined grammar, LLMs can interpret queries that include dialect-specific features. \sqlflex{} first attempts to parse a query using the grammar-based parser. When parsing fails due to dialect-specific features, \sqlflex{} invokes the segmenter to split the input into smaller parts such that they can be parsed or further segmented. This decomposes hierarchical parsing into sequential segmentation tasks that better align with the strengths of LLMs. Since clause-level grammar is typically flat, while expressions are recursive, \sqlflex{} employs clause-level segmentation and expression-level segmentation as two strategies suited to these different syntactic forms. Additionally, to improve reliability, \sqlflex{} validates the output by checking properties between the input and the output after each segmentation.

We conducted a large-scale evaluation of \sqlflex{}. Specifically, we showed \sqlflex{}'s practicality in two real-world use cases, SQL linting and test-case reduction, each evaluated on a task-specific dataset. For linting, \sqlflex{} surpasses SQLFluff in ANSI mode by 63.68\% in F1 score and performs on par with SQLFluff in dialect-specific mode (98.14\% vs.\ 98.24\%). For test-case reduction, \sqlflex{} outperforms \sqless{} by up to 10 times in simplification rate. Furthermore, we evaluated \sqlflex{}'s dialect-agnostic parsing effectiveness, using queries in eight different SQL dialects extracted from their respective DBMS test suites. \sqlflex{} successfully parsed 91.55\% to 100\% of queries across the eight dialects, outperforming both a PostgreSQL-specific parser and \sqlglot{}. On the most challenging dialect, \sqlflex{} achieves improvements of up to 179.46\% over the PostgreSQL-specific parser and 138.04\% over \sqlglot{}.

We believe that \sqlflex{} could be the foundation for many SQL-related use cases, where no manual effort is needed to support parsing of dialect-specific features. Like many other LLM-based applications (\eg{} Text-to-SQL~\cite{li2024DawnNaturalLanguage, liCanLLMAlready}), \sqlflex{} relies on a best-effort approach, so we defer exploring correctness-critical use cases such as SQL-level query optimization~\cite{wangWeTuneAutomaticDiscovery2022, zhou2021SIAOptimizingQueries, baiQueryBoosterImprovingSQL2023} as part of future work. To strengthen correctness guarantees, \sqlflex{} could potentially be integrated with query equivalence verification tools~\cite{verieql24he, wang24QED}.


To summarize, we make the following contributions:\footnote{Our artifact is publicly available at https://github.com/wanteatfruit/SQLFlex and https://doi.org/10.5281/zenodo.18975512.} 
\begin{itemize}[leftmargin=*]
    \item We propose the novel idea of integrating the strengths of grammar-based parsers and LLMs for query parsing.
    \item We propose \sqlflex{}, a dialect-agnostic query rewriting framework which implements this idea with strategies for clause-level and expression-level segmentation.
    \item We extensively evaluated \sqlflex{} in two real-world use cases and in a standalone evaluation.
\end{itemize}

\section{Background and Motivation}\label{sec:background}

\begin{lstlisting}[style=tsql, label={lst:walkthrough}, caption={TSQL query, dialect-specific features highlighted}, float]
SELECT @TOP 10@ *
FROM Sales
WHERE (tot / 2) @!<@ 8 AND @year@ < 2025
@OPTION (FAST 10)@
\end{lstlisting}

\begin{lstlisting}[style=custom, caption={Simplified ANTLR grammar of a query}, label={lst:ebnf-query}, float]
selectStmt: selectClause fromClause? whereClause?;
selectClause: "SELECT" selectSpec? projections;
selectSpec: "DISTINCT" | "ALL";
fromClause: "FROM" tableReferences;
whereClause: "WHERE" expr;
expr: expr op expr | identifier | number;
op: "AND" | "<" | "/";
\end{lstlisting}

\paragraph{SQL standard and dialects}
SQL is a standardized declarative language for data manipulation in RDBMSs~\cite{ansi-sql}. We use \emph{SQL standard} to refer specifically to SQL-92~\cite{sql92}, a minimal version \emph{``supported by virtually all RBMSs''}~\cite{critiqueOfModern24thomas}. The core operation in SQL is the query, typically the \codeIn{SELECT} statement. Elements in a query can be grouped into \emph{clauses} and \emph{expressions}. A query is composed of multiple clauses, each of which specifies a particular aspect of the query. Clauses contain \emph{clause elements}, which include both expressions and non-expression elements that contribute to the clause’s behavior. Expressions are composable units made up of values, operators, functions, or subqueries, and they evaluate to a result~\cite{duckdb-expressions}. Listing~\ref{lst:walkthrough} shows an example query in the TSQL dialect~\cite{TSQL}, which contains the \codeIn{SELECT}, \codeIn{FROM}, \codeIn{WHERE}, and \codeIn{OPTION} clauses, where each clause includes the keyword and its associated elements. For example, within the \codeIn{SELECT} clause, \codeIn{``TOP 10''} is a non-expression clause element that limits the number of returned rows, and \codeIn{``*''} is a wildcard expression selecting all columns. Similarly, the \codeIn{WHERE} clause contains a predicate expression as a clause element that filters the result set.

In practice, DBMSs implement their own variants of SQL, known as \emph{SQL dialects}~\cite{critiqueOfModern24thomas}. For example, SQL Server uses the TSQL dialect. Although the dialects share a common foundation (\eg{} SQL-92), they differ in grammar rules, reserved keywords, and DBMS-specific features~\cite{zhou2025cracking}. For instance, unlike DuckDB, MySQL requires an alias for subqueries. In another case, \codeIn{``OPTION''} has no special meaning in PostgreSQL, but in TSQL it is a reserved keyword and must be quoted when used as an identifier. A recent study found that test suites contain mostly dialect-specific features, where approximately 70\% of queries in the PostgreSQL test suite are incompatible with other DBMSs, and about 60\% for DuckDB~\cite{zhong2024UnderstandingReusingTest}. In this paper, we focus on \codeIn{SELECT} statements, and refer to syntax unsupported by the SQL standard as \emph{dialect-specific features}.

\paragraph{Grammar-based parsers} 
Grammar-based parsers build an AST from an input query using a formal grammar. ANTLR~\cite{antlr} is a widely used parser generator that takes a grammar specification and generates a parser reflecting the grammar’s structure. A grammar consists of \emph{production rules} that define how symbols can be expanded. A \emph{symbol} refers to any element in a production rule, which is either a \emph{terminal} or a \emph{non-terminal}. Listing~\ref{lst:ebnf-query} illustrates a simplified ANTLR grammar for a \codeIn{SELECT} statement in the SQL standard. It denotes \emph{terminals} in quotes and defines \emph{non-terminals} using rule names. For instance, the rule \codeIn{selectClause} includes the terminal \codeIn{"SELECT"} and two non-terminals. An important observation from the SQL standard grammar is that clauses (\eg{} \codeIn{selectClause}) tend to be non-recursive, while expressions are typically recursive.

\paragraph{Existing approaches}
Most query rewriting tools rely on grammar-based parsers to construct an AST, but these parsers often fail on dialect-specific features, preventing further rewriting. We categorize grammar-based parsers as dialect-specific and multi-dialect parsers. For dialect-specific parsers, although they work well for the supported dialect, they lack generalizability across other dialects. Additionally, not all dialects have a ready-to-use parser in practice. This limitation also applies to parser generators like ANTLR, as grammars for many dialects are unavailable or incomplete~\cite{antlr-mysql-grammar-issue}. In contrast, multi-dialect parsers aim to produce unified ASTs across dialects, offering better generalizability. Notable examples include the parser component in \sqlglot{}~\cite{SQLGlot}, \sqlfluff{}~\cite{SQLFluff}, and Calcite~\cite{Calcite}. While effective on multiple dialects, these tools demand substantial human effort and expertise to maintain the codebase and support additional dialects. For instance, the entire parser in \sqlglot{} contains over 20,000 lines of code, and adding support for an additional dialect often requires more than 1,000 additional lines. This high implementation cost makes such approaches difficult to scale, as evidenced by the many unresolved dialect-related feature requests in their open-source repositories~\cite{calcite-doris-issue, sqlfluff-h2-issue, sqlglot-dremio-issue}.

\begin{figure*}
    \centering
    \includegraphics[width=\linewidth]{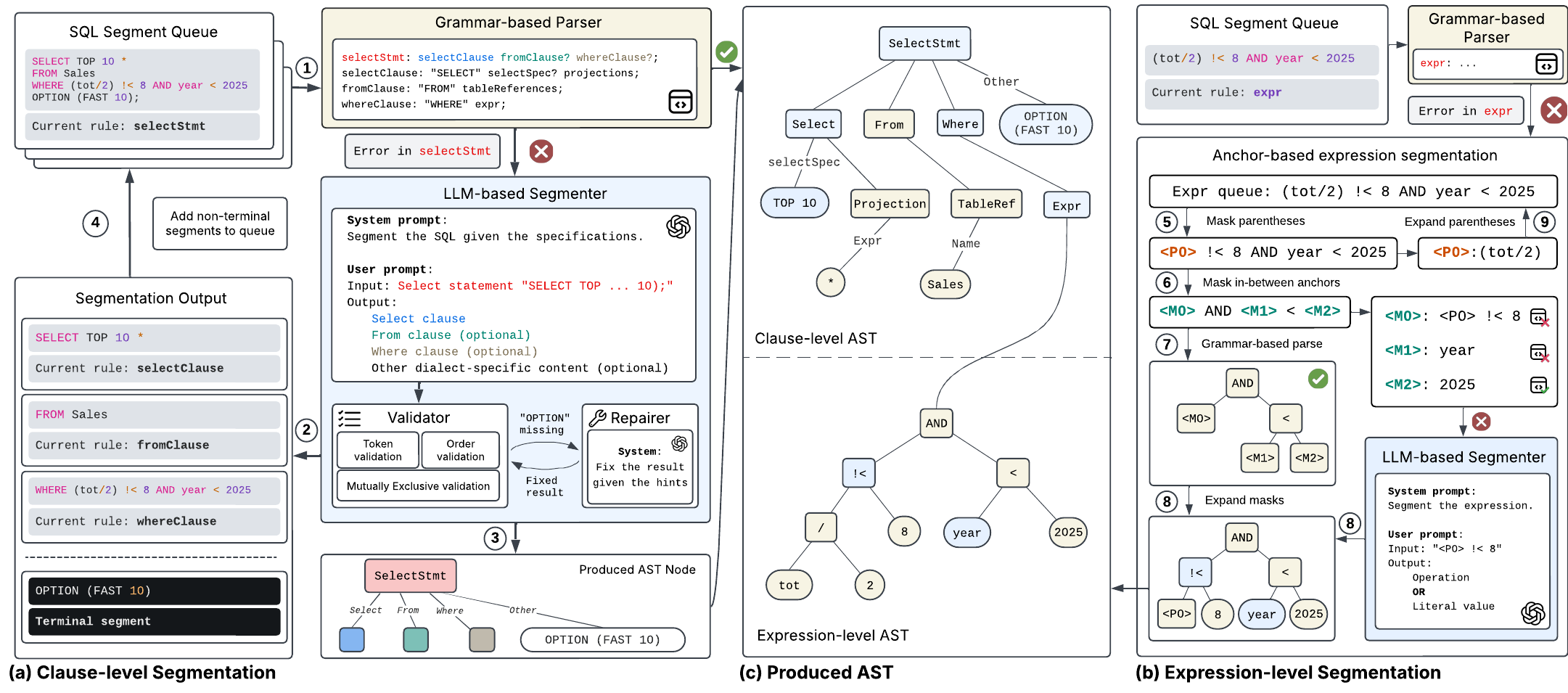}
    \caption{Illustrative example of hybrid segmentation. Background colors of nodes denote their origins; text colors map grammar rules with segmentation prompts in (a).}
    \Description{Illustration of the segmenter process. }
    \label{fig:walkthrough}
\end{figure*}

\section{Illustrative Example}
We use the TSQL query in Listing~\ref{lst:walkthrough} to illustrate the challenges of dialect-agnostic query parsing and how \sqlflex{} constructs its AST via hybrid segmentation. Dialect-specific features in the listing are colored, including the \codeIn{OPTION} clause for query hints, the \codeIn{``TOP 10''} select specifier, the not-less-than \codeIn{``!<''} operator, and the use of \codeIn{``year''} as an identifier despite being a reserved keyword in standard SQL. Grammar-based parsing alone fails here, as these features fall outside the grammar rules.

The core of our approach is \emph{segmentation}, a divide-and-conquer strategy for handling dialect-specific features. When such features cause the grammar-based parser to fail, we prompt an LLM to decompose the query into multiple segments---aligning with the LLM’s strengths in sequential processing~\cite{jiang2025hibenchbenchmarkingllmscapability, segmentAnyText2024}---and iteratively process each segment using our hybrid approach. For example, when parsing the query in Listing~\ref{lst:walkthrough}, the grammar-based parser would fail, as \codeIn{``TOP''} is an invalid terminal symbol in the lower-level \codeIn{selectSpec} grammar rule. While it might be clear that \codeIn{``TOP 10''} should be attributed to the \codeIn{SELECT} clause rather than the \codeIn{FROM} clause, as the \codeIn{SELECT} clause would be incomplete otherwise, deciding so is difficult due to the English-like syntax of SQL~\cite{critiqueOfModern24thomas, sqlHasProblems24Shute}. Specifically, consider the \codeIn{OPTION} keyword in Listing~\ref{lst:walkthrough}. In PostgreSQL, \codeIn{OPTION} is accepted as an identifier, whereas in TSQL it marks the beginning of an \codeIn{OPTION} clause. Without dialect knowledge, it is ambiguous whether \codeIn{OPTION} should be interpreted as part of the \codeIn{WHERE} clause or as a separate clause parallel to \codeIn{WHERE}. Thus, rather than relying on heuristics, we use a state-of-the-art LLM that has been trained on examples of various SQL dialects to obtain more accurate parsing outputs.

Figure~\ref{fig:walkthrough} illustrates our segmentation approach. When parsing the full query fails (\ding{172}), we prompt the LLM to segment the query into three segments corresponding to the three non-terminals \codeIn{selectClause}, \codeIn{fromClause}, and \codeIn{whereClause}, as well as dialect-specific elements that correspond to neither of the non-terminals (\ding{173}). To enable this, we map non-terminals to prompts, in addition to a dedicated prompt for dialect-specific segments, each instructing the LLM to extract the corresponding segment from the input. These mappings are visualized using text colors in the figure. In the example, the returned segments are \codeIn{``SELECT TOP 10 *''} for \codeIn{selectClause}, \codeIn{``FROM Sales''} for \codeIn{fromClause}, \codeIn{``WHERE}\dots\codeIn{2025''} for \codeIn{whereClause}, and \codeIn{``OPTION (FAST 10)''}, the latter of which is a dialect-specific segment that cannot be further parsed. After segmentation, we create the corresponding \codeIn{SelectStmt} node (\ding{174}) to build the AST top-down. The parsing process continues on the parseable segments (\ding{175}). For example, in a subsequent iteration, the LLM segments the \codeIn{SELECT} clause into \codeIn{``TOP 10''} and \codeIn{``*''}, mapping them to \codeIn{selectSpec} and \codeIn{projections}, respectively. Since the \codeIn{FROM} clause lacks dialect-specific features, grammar-based parsing is sufficient. We refer to this approach as \emph{clause-level segmentation}, which we detail in Section~\ref{sec:clause-level}. This approach is effective for non-recursive structures, which are present at the clause level.

At the expression level, clause-level segmentation is error-prone for deeply nested expressions. While most state-of-the-art LLMs perform well on simple cases, we observe that complex expressions are likely to be misinterpreted. For example, let \codeIn{e} be the \codeIn{WHERE} clause expression in Listing~\ref{lst:walkthrough}. When we wrap \codeIn{e} inside a \codeIn{CASE} expression (\eg{} \codeIn{CASE WHEN e THEN e ELSE e END}), the LLM may misinterpret the \codeIn{``AND''} operator in \codeIn{e} as having the lowest precedence, overlooking the \codeIn{CASE} expression. Additionally, applying clause-level segmentation to such expressions requires invoking the LLM at every level of the expression tree, causing a high computational overhead. To address this, our key observation is that certain operations, such as \codeIn{``AND''} and \codeIn{``<''}, are recognized by the grammar, making their relative positions in the AST clear even in the presence of dialect-specific features. Additionally, parenthesized subexpressions (\eg{} \codeIn{(tot/2)}) indicate higher precedence and appear deeper in the AST, so we can defer their handling until the outer parts have been parsed. Motivated by these observations, we propose \emph{expression-level segmentation}, a strategy designed to handle the expressions.

The core idea of \emph{expression-level segmentation} is to treat known operators as \emph{anchors} and replace fragments between them with \emph{abstraction tokens}. This enables the grammar-based parser to parse the anchors and isolate the dialect-specific features. Figure~\ref{fig:walkthrough}b illustrates one iteration of expression-level segmentation, which begins when a query segment fails to be parsed using the expression-level \codeIn{expr} rule. First, all parenthesized subexpressions are replaced with \emph{parenthesis tokens} (\eg{} \codeIn{(tot/2)} becomes \codeIn{<P0>}) so that the current iteration focuses only on the non-parenthesized part (\ding{176}). Next, in the remaining expression, recognizable operators (\ie{} anchors) \codeIn{``AND''} and \codeIn{``<''} are preserved, while the fragments between them are replaced with \emph{mask tokens} (\eg{} \codeIn{<P0> !< 8} becomes \codeIn{<M0>}) (\ding{177}). The abstracted expression \codeIn{``<M0> AND <M1> < <M2>''} can now be parsed by the grammar, as dialect-specific features are isolated within mask tokens (\ding{178}). After obtaining the AST of the abstracted expression, we expand each mask token (\ding{179}), allowing the LLM to process only the fragments containing dialect-specific features rather than the full expression. Specifically, if grammar-based parsing fails for a token, we perform segmentation. For example, \codeIn{<M0>} and \codeIn{<M1>} cannot be parsed due to the dialect-specific \codeIn{!<} operator and reserved keyword \codeIn{year} used as an identifier. The segmenter identifies \codeIn{<M0>} as an operation with \codeIn{``!<''} as the operator and \codeIn{``<P0>''} and \codeIn{``8''} as operands. Once the dialect-specific operator is segmented, the operands can be parsed separately. For \codeIn{<M1>}, the segmenter treats \codeIn{``year''} as a literal, indicating that it is already a terminal node. \codeIn{<M2>} is valid and parsed directly. Lastly, we expand the parenthesis tokens and add the enclosed expressions to the next iteration (\ding{180}). For instance, \codeIn{``tot/2''} in \codeIn{<P0>} is processed in the next iteration. It contains no dialect-specific feature and is successfully parsed. We detail \emph{expression-level segmentation} in Section~\ref{sec:expr-level}.

Figure~\ref{fig:walkthrough}c shows the resulting AST after hybrid segmentation, where non-terminal nodes are visualized as rectangles and terminal nodes as ovals. Background colors indicate whether a node was produced by the grammar-based parser or the LLM-based segmenter. To improve reliability and reduce hallucinations, we apply validation and repair mechanisms to the segmentation outputs. The next section describes the overall approach in detail.

\section{SQLFlex}

We present \sqlflex{}, a dialect-agnostic query rewriting framework. We focus the rest of this section on detailing our key contribution, the hybrid segmentation process. As both AST manipulation and pretty-printing are based on standard techniques, we provide only a high-level overview of them in Section~\ref{ref:rewriter}.

Algorithm~\ref{alg:segmentation-overall} outlines the hybrid segmentation process, which builds the AST iteratively from top to bottom. Formally, given a query fragment $Q$, we define \emph{segmentation} as a function $S: Q \rightarrow {s_1, s_2, \dots, s_n}$, where each $s_i$ is an output segment. We maintain a queue to manage query fragments whose sub-ASTs still need to be built; each entry is a tuple $(Q, G)$, where $G$ is the grammar rule used to parse $Q$. We initialize the queue with the query and \codeIn{selectStmt} rule. In each iteration, we dequeue one item from the queue. Since grammar-based parsing is more efficient and reliable, we first attempt to parse $Q$ using $G$ (line 5). If parsing fails, the algorithm falls back on clause-level or expression-level segmentation depending on $G$ (lines 7--10). Expression-level segmentation is used only when $G$ is \codeIn{expr}, indicating the parsing of an expression element; in all other cases, clause-level segmentation applies. Since clause-level segmentation produces intermediate segments that require further parsing, we add them to the queue. Both expression-level segmentation and grammar-based parsing construct the full sub-AST, so no additional fragments are enqueued.

\SetKwFor{While}{while}{do}{}
\SetKwInput{KwInput}{Input}
\SetKwInput{KwResult}{Result}
\SetKw{KwReturn}{return}
\SetAlCapFnt{\footnotesize}
\SetKwComment{Comment}{$\triangleright$\ }{}
\SetCommentSty{mycommfont}
\DontPrintSemicolon
\begin{algorithm}[t]
\footnotesize
\setstretch{1.05}
\caption{\footnotesize Hybrid Segmentation}
\label{alg:segmentation-overall}
\KwInput{\emph{input\_query} \Comment*[r]{Full input query}}
\KwResult{\emph{root} \Comment*[r]{Parsed query AST}}

\SetKwProg{Fn}{Function}{:}{}

\Fn{HybridSegmentation}{
\emph{seg\_queue} $\gets$ \emph{[(input\_query, selectStmt)]}, \emph{root} $\gets$ $\emptyset$\;

\While{seg\_queue not empty}{
    \emph{Q, G} $\gets$ \emph{seg\_queue.dequeue()} \;
    \uIf(\Comment*[f]{Grammar-based parsing}){parse(Q, G) succeeds}{
        \emph{ast} $\gets$ \emph{parse(Q, G)}
    }
    \uElseIf{G == expr}{
        \emph{ast} $\gets$ \emph{ExpressionLevelSegmentation(Q)}
    }
    \Else{
        \emph{ast} $\gets$ \emph{ClauseLevelSegmentation(Q, G, seg\_queue)}
    }
    \emph{root} $\gets$ \emph{attach(root, ast)}\;
}
\KwReturn{root}\;
}
\Fn{ClauseLevelSegmentation(Q, G, seg\_queue)}{
    \emph{rule2ast, rule2prompt, segment2rule} $\gets$ \emph{load\_mapping()}\;
    \emph{N} $\gets$ \emph{rule2ast[G]} \Comment*[r]{AST node type for $G$}
    \emph{prompts} $\gets$ \emph{rule2prompt[G]} \Comment*[r]{Prompts associated with $G$}
    \emph{segments} $\gets$ \emph{Segmentation(Q, prompts)}\;
    \For{$s_i$ $\in$ segments}{
        \emph{$g_i$} $\gets$ \emph{segment2rule[$s_i$]} \Comment*[r]{Maps each segment to one $g_i$}
        \uIf{has\_nonterminal($g_i$)}{
            \emph{seg\_queue.enqueue(($s_i$, $g_i$))}\;
        }
        \Else{
            \emph{N} $\gets$ \emph{attach(N, $s_i$)}\;
        }
    }
    \KwReturn{N}\;
}


\end{algorithm}

    


\subsection{Clause-level Segmentation}\label{sec:clause-level}
The core idea of clause-level segmentation is to map each non-dialect-specific segment $s_i$ to a corresponding symbol $g_i$ in the grammar rule $G$, so that $s_i$ can be further parsed or segmented using $g_i$ in subsequent iterations. Segments that lack a symbol mapping are treated as terminals. The clause-level segmentation function is also outlined in Algorithm~\ref{alg:segmentation-overall}, which takes as input the current query fragment $Q$, the associated grammar rule $G$, and the shared queue from the hybrid segmentation loop.

To support clause-level segmentation, we predefine three mappings (line 14). First, \emph{rule2ast} maps each grammar rule $G$ to its corresponding AST node $N$ (line 15). For instance, \codeIn{selectStmt} maps to a \codeIn{SelectStmt} node. This information is needed to instantiate the correct AST node. Second, \emph{rule2prompt} maps $G$ to segmentation prompts, each associated with a symbol $g_i$ (line 16). For example, \codeIn{selectStmt} includes the \codeIn{selectClause} symbol, mapping to a prompt \emph{``Output should include a SELECT clause''}. We use few-shot prompting with examples of common SQL features to improve consistency and reduce bias toward a specific dialect. The prompts remain unchanged when applying \sqlflex{} across dialects. Lastly, \emph{segment2rule} maps each output segment $s_i$ back to its corresponding $g_i$, forming $(s_i, g_i)$ tuples for the queue (line 19). For example, a segment corresponding to a \codeIn{SELECT} clause will be mapped to \codeIn{selectClause}. This helps the approach to resume parsing segments with the correct grammar rule. 




After segmentation, each segment is processed based on the mapped $g_i$ to determine whether it will be enqueued (lines 18--23). If $g_i$ contains non-terminal symbols (\eg{} \codeIn{selectClause} contains non-terminal \codeIn{selectSpec}), the $(s_i, g_i)$ tuple is added to the queue (lines 20--21). Dialect-specific features within these segments are handled in subsequent iterations to maintain sequential processing. In contrast, if $g_i$ contains only terminal symbols, the segment is instead directly attached to the created AST node $N$ as no further parsing is required (lines 22--23). This avoids repeated failures caused by dialect-specific features mapped to such rules. For example, in an iteration where $Q$ is \codeIn{``SELECT TOP 10 *''} and $G$ is \codeIn{selectClause}, the segment \codeIn{``TOP 10''} maps to the terminal-only rule \codeIn{selectSpec}, and is therefore attached to $N$ directly without being enqueued. A special case occurs when a segment represents a dialect-specific feature lacking a corresponding $g_i$, such as the \codeIn{OPTION} clause. These segments are encapsulated in dedicated \codeIn{Other} terminal nodes and attached to $N$. We record their position in the query for correct pretty-printing.  Section~\ref{sec:use-case} shows that the core grammar suffices for various rewriting tasks, while content in \codeIn{Other} nodes can be handled with text-based analysis. Our approach outperforms heuristic-based methods, as it preserves the structure of the AST for these nodes (\eg{} the \codeIn{OPTION} clause is placed at the same level as the other clauses).

Clause-level segmentation returns the constructed AST node to the main hybrid segmentation loop (line 26). In subsequent iterations, the items added to the queue during clause-level segmentation are processed. If a dequeued item is $G = expr$ and parsing fails, the algorithm invokes expression-level segmentation.

\subsection{Expression-level Segmentation}\label{sec:expr-level}
We propose an anchor-based strategy to isolate dialect-specific features into expression fragments that can be independently parsed. Specifically, the strategy leverages the precedence of parentheses and known operators (\eg{} \codeIn{AND}) to construct partial ASTs, so that segmentation can be performed directly on their operands. We refer to it as \emph{anchor-based expression segmentation}, outlined in Algorithm~\ref{alg:expr}. The algorithm takes an expression as input and builds its AST in a top-down iterative manner. A queue manages the expression fragments to be processed, where each iteration processes one level of parentheses, starting with the input expression.


\SetKwFor{While}{while}{do}{}
\SetKwInput{KwInput}{Input}
\SetKwInput{KwResult}{Result}
\SetKw{KwReturn}{return}
\SetAlCapFnt{\footnotesize}
\SetKwComment{Comment}{$\triangleright$\ }{}
\SetCommentSty{mycommfont}
\DontPrintSemicolon
\begin{algorithm}[t]
\footnotesize
\setstretch{1.05}
\caption{\footnotesize Anchor-based Expression Segmentation}
\label{alg:expr}
\KwInput{\emph{input\_expr} \Comment*[r]{Full input expression}}
\KwResult{\emph{expr\_root} \Comment*[r]{Expression-level AST}}

\SetKwProg{Fn}{Function}{:}{}

\Fn{ExpressionLevelSegmentation(input\_expr)}{
    \emph{expr\_queue} $\gets$ \emph{[input\_expr]},  \emph{expr\_root} $\gets$ $\emptyset$\;
    \emph{prompts, anchors} $\gets$ \emph{load\_prompts\_and\_anchors()} \Comment*[r]{Predefined}
    \While{expr\_queue not empty}{
        \emph{$E$} $\gets$ \emph{expr\_queue.dequeue()}\;
        \emph{$E'$}, \emph{paren\_map} $\gets$ \emph{process\_paren}(\emph{$E$})\;
        \emph{$E''$}, \emph{mask\_map} $\gets$ \emph{process\_anchor}(\emph{$E'$, anchors})\;
        \emph{expr\_root} $\gets$ \emph{parse\_and\_attach}(\emph{$E''$}, \emph{expr\_root}) \Comment*[r]{Produces $T$}
        \For{$m_i$, $e_i$ $\in$ mask\_map}{
            \uIf{parse($e_i$) succeeds}{
                \emph{expr\_root} $\gets$ \emph{attach(parse($e_i$), expr\_root)}\;
            }
            \Else{
                \emph{expr\_root} $\gets$ \emph{attach(RecursiveSeg($e_i$, prompts), expr\_root)}\;
            }
        }
        \For{$p_i$, paren\_content $\in$ paren\_map}{
            \emph{expr\_queue.enqueue(remove\_paren(paren\_content))}\;
        }
    }
\Return{expr\_root}\;
}

\end{algorithm}

Given an expression $E$ and a set of \emph{anchors} (\ie{} known operators), the isolation process builds a partial AST $T$, where each leaf node contains a fragment $e_i$ that may contain dialect-specific features. This divide-and-conquer approach allows each fragment to be segmented independently, preventing the LLM from misinterpreting nested subexpressions. This process consists of three steps, which we illustrate using the same expression in Figure~\ref{fig:walkthrough}b as $E$. First, we replace each parenthesized subexpression with a unique \emph{parenthesis token} $p_i$, returning the transformed expression $E'$ and mappings from each $p_i$ to its original content (line 6). For example, \codeIn{(tot/2)} in $E$ is replaced with $p_0$, and we record the mapping $p_0 \mapsto$ \codeIn{(tot/2)}. Next, we identify anchors in $E'$ and replace surrounding content with \emph{mask tokens} $m_i$, returning the transformed $E''$ and mappings from each $m_i$ to its corresponding fragment $e_i$ (line 7). For example, the anchors in $E'$ are \codeIn{``AND''} and \codeIn{``<''}, thus $E'$ can be split into fragments $e_i$ and anchors $a_i$ as ``$e_0 a_0 e_1 a_1 e_2$'', where $a_0$ is \codeIn{``AND''}, $e_0$ is \codeIn{``p0 !< 8''}, etc. We replace each $e_i$ with a mask token $m_i$ and record their mappings. Lastly, we extend the base grammar to accept both token types, so that the masked expression $E''$ is parseable by the grammar-based parser. The produced $T$ preserves the precedence of known operators (line 8). For dialect-specific operators whose precedence is unknown, our approach assumes they have the highest precedence among all unparenthesized operators, as low-precedence operators are typically reserved for common operators like \codeIn{``AND''} and \codeIn{``OR''} to compose multiple expressions. Determining the exact operator precedence is non-trivial both for the LLM and for users, as dialect-specific precedence rules are often inconsistent across DBMSs. In practice, we expect users to disambiguate the precedence of dialect-specific features that have a low precedence by adding parentheses, making operator precedence explicit, which is a best practice suggested by style guides~\cite{sqlstyleguide}. 





We process each expression fragment $e_i$. For each masked fragment, the algorithm first attempts grammar-based parsing. Some fragments can be successfully parsed, as they may consist of recognized literal values or parenthesis tokens (including function calls where their argument lists are replaced by $p_i$). If parsing succeeds, the resulting subtree is attached directly to tree $T$ (lines 10--11). If parsing fails, we recursively invoke the segmenter to handle the dialect-specific content and attach the resulting AST (lines 12--13).




At its core, the segmenter distinguishes between terminal and non-terminal fragments. Terminal fragments require no further parsing, while non-terminal fragments must recursively parse their child expressions to complete the structure. Thus, we prompt the LLM to output \emph{either} a literal value segment or operation segments (\ie{} an operator and its operands). Literal segments are treated as terminal nodes (\eg{} \codeIn{year}). For operation segments, the LLM identifies the operator with the lowest precedence (\ie{} the shallowest AST node) and its operands. Each operand is then recursively parsed or segmented. For example, given the fragment \codeIn{``<P0> !< 8''}, the LLM identifies \codeIn{!<} as an operator, with operands \codeIn{<P0>} and \codeIn{8}, both of which can be parsed directly and attached to the tree. In more complex cases, such as \codeIn{``1 !< 2 !< 3''}, the initial segmentation identifies the first \codeIn{!<} as a left-associative operator, with left operand \codeIn{``1''} and right operand \codeIn{``2 !< 3''}. Since the right operand still contains a dialect-specific operator, segmentation is invoked again, continuing recursively until all terminal nodes are resolved.

After processing all fragments $e_i$, we handle the parenthesized expressions. For each such expression (\emph{paren\_content}), the outermost pair of parentheses is removed. The content is treated as a single expression and added to the queue for further expression-level segmentation (line 15). However, in two special cases, we proceed differently. First, if the content begins with \codeIn{``SELECT''}, we perform clause-level segmentation for the subquery. Second, if the content has unparenthesized commas---common in function arguments---it is treated as an expression list and split into separate expressions, each enqueued individually. For example, given the function \codeIn{``max(1, min(a, 1))''}, we process its argument list. We remove the outer parentheses and split the arguments into \codeIn{1} and \codeIn{min(a, 1)}, enqueuing both. The comma in \codeIn{min(a, 1)} remains parenthesized and will be handled in subsequent iterations. The algorithm continues until all parentheses are resolved (\ie{} no more expressions in the queue), and returns the expression-level AST. After combining with the clause-level AST, this yields the full AST for the input query.

\subsection{Validation and Repair}\label{sec:val}
For each segmentation output, regardless of the strategy, we apply three validation methods that check invariant properties between the input and the segmented output to detect potential errors. The LLM is prompted to repair its output until all validation checks pass. Note that while this approach improves robustness, it remains best-effort, as it might miss mistakes made by the LLM.
\begin{enumerate}[leftmargin=*]
    \item \emph{Token validation.} We compare the characters in each segment with those in the input to detect any missing or extraneous characters. For example, if \codeIn{``OPTION''} is omitted during segmentation of the entire query, the validation fails. 
    
    \item \emph{Order validation.} We ensure that the character order within each segment matches the order in the original input. For example, if \codeIn{``FAST 10''} is reordered as \codeIn{``10 FAST''}, the validation fails. We also check the semantic order among the segments. For example, in expression-level segmentation, if the left operand appears after the operator, the validation also fails.
    \item \emph{Mutually exclusive output validation.} Expression-level segmentation defines mutually exclusive output types where the output is either a literal or an operation. If both output types are present, the validation fails.
    
    

    

\end{enumerate}

Repair prompts are created based on the failed validation, the reason for the failure, and the full query to provide more context. After each repair attempt, the output is revalidated, repeating until success or reaching a predefined retry limit. If the limit is reached, we attach a special \codeIn{Unsegmented} terminal node to the AST. This node preserves the original input and prevents further processing of that segment to maintain the overall structure of the AST. For example, if the \codeIn{SELECT} clause fails to be segmented, an \codeIn{Unsegmented} node containing the clause is attached to the \codeIn{SelectStmt} node. The rest of the query (\eg{} \codeIn{FROM} clause) can still be processed normally.

\subsection{Rewriter and Pretty-Printer}\label{ref:rewriter}
While the \emph{Rewriter} modifies the AST based on user-provided logic, the \printer{} regenerates the query from the AST. We consider neither of them novel and describe them only for completeness.

For the \emph{Rewriter}, we provide two APIs to users: \codeIn{find} and \codeIn{transform}. The transform function is a higher-order API that applies a user-defined rule to modify AST nodes. It traverses the AST and applies the rule to each node. The \codeIn{find} function complements this by searching for nodes of a specified type and returns nodes that satisfy an optional filter function. We designed the API to be minimal to support flexible and dialect-agnostic rewriting. Listing~\ref{lst:rewrite-example} shows an example of how users can detect and remove an unused table alias given an input query using \sqlflex{}. More complex AST manipulation can be implemented similarly by composing calls to \codeIn{find} and \codeIn{transform}, along with more extensive user-defined functions. Users can also incorporate dialect-specific knowledge to directly manipulate node attributes, including those representing dialect-specific features.

The \printer{} converts an AST back into its corresponding SQL query, ensuring that any modifications made by the \rewriter{} are reflected in the final output. It is a purely rule-based system that traverses the AST to regenerate the source text~\cite{cooper2003EngineeringCompiler}.

\begin{lstlisting}[style=python, caption = {Detect and remove unused table alias}, label = {lst:rewrite-example}, float]
# detect anti-pattern using find API
def find_unused_table_alias(root):
  def is_table_alias(node): # filter function
    return isinstance(node.parent, Table)
  aliases = find(Alias, root, is_table_alias)
  identifiers = find(Identifier, root)
  if al in aliases and not in identifiers:
    return al

# user-defined rewrite rule for transform API
def rmv_table_alias(node, al):
  if isinstance(node, Table) and node.alias == al:
    node.alias = None

root = sqlflex_parse(query)
unused = find_unused_table_alias(root)
new = transform(root, rmv_table_alias, unused)
\end{lstlisting}
\section{Implementation and Experimental Setup}\label{sec:impl}
We implemented \sqlflex{} in around 3,800 lines of Python. We adopted the \sqlninetytwo{}~\cite{sql92} grammar as the base grammar. It has a minimal grammar that is supported by almost all SQL dialects~\cite{critiqueOfModern24thomas}, while allowing us to highlight the effectiveness of segmentation. We used ANTLR~\cite{antlr} to implement the grammar-based parser. We implemented the LLM‑based segmenter with LangChain~\cite{langchain} and the OpenAI GPT‑4.1 API~\cite{gpt4.1}, run at zero temperature for more deterministic outputs. We use a system prompt defining the general segmentation task (Listing~\ref{lst:sys_prompt}) and allow up to three repair attempts to balance runtime and success rate. We used the LLM's structured output feature to ensure its outputs align with the prompts and can be subsequently processed. The output structure is defined using \codeIn{pydantic}. For example, Listing~\ref{lst:prompt} shows the structured output prompt for segmenting the \codeIn{selectStmt} rule, where the \codeIn{pydantic} model maps to the \codeIn{SelectStmt} AST node definition. The \codeIn{SelectStmt} class represents the input node type, and its attributes correspond to the expected output fields. Each field is annotated with a natural language description, which is automatically incorporated into the LLM prompt to guide structured generation. {Note that the manual effort required to define these descriptions is incurred only once. We used LLMs to assist in generating and refining the descriptions, with the overall process taking less than ten minutes.}

We ran all subsequent experiments on a server with a 64-core AMD EPYC 7763 processor (2.45GHz) and 512GB of RAM, running Ubuntu 22.04.

\begin{lstlisting}[style=boxed, caption={System prompt for LLM-based segmenter}, label={lst:sys_prompt}, float]
You are an SQL expert in various SQL dialects. Your task is to segment an SQL query or a valid part of a query into its components in the {dialect} SQL dialect while adhering strictly to the following rules:
- Ensure all tokens from the original query are included in the segmented output. Do not add, modify, or omit any tokens.
- Maintain the relative order of tokens as they appear in the input.

Important Notes:
- The input is always a valid SQL query or a valid fragment of a query in {dialect}.
- Use context information to understand the query.
- Reason step by step to ensure correctness and consistency: self-reflect on 3 to 5 results, and return the most consistent result as the final output.

\end{lstlisting}

\begin{lstlisting}[style=python_prompt, caption={Structured output prompt fragment for \codeIn{selectStmt}}, label={lst:prompt}, float]
class SelectStmt(BaseModel):
  """Represents a SQL SELECT statement."""
  select: str = Field(description="The SELECT clause, including the **SELECT** keyword and its associated projections.")
  from_: Optional[str] = Field(description="The FROM clause, including the **FROM** keyword, specifying the source table(s).")
  where: Optional[str] = Field(description="The WHERE clause, including the **WHERE** keyword, defining filtering conditions.")
  other: Optional[str] = Field(description="Any additional dialect-specific contents.")
\end{lstlisting}

\section{Use Cases}\label{sec:use-case}

We present two real-world query analysis and rewriting use cases implemented using \sqlflex{} as the base parser, namely SQL linting and test-case reduction. SQL linting is a query analysis task, while test-case reduction involves query rewriting and requires executing the queries. The use cases highlight \sqlflex{}’s practical utility, which grounds our evaluation in actual usage of \sqlflex{} before turning to a standalone evaluation.

\subsection{SQL Linting}
Many SQL queries exhibit poor coding practices that could cause performance and maintainability issues~\cite{zhu2022EmpiricalStudyQuality, smellyRelations2018}, commonly referred to as SQL \emph{anti-patterns}~\cite{sqlAntiPatterns2010}. A prevalent example is the usage of the column wildcard expression (\eg{} \codeIn{SELECT *}) to retrieve all columns from a table. While this shorthand is convenient, it is discouraged in production, as it can lead to inefficiencies and cause schema changes to remain undetected~\cite{sqlAntiPatterns2010}. Hence, a robust SQL linter is essential for detecting anti-patterns and improving code quality~\cite{SQLFluff, dintyalaSQLCheckAutomatedDetection2020, smellyRelations2018}.

\paragraph{Baselines}
We chose two notable existing linters, \sqlcheck{}~\cite{dintyalaSQLCheckAutomatedDetection2020} and \sqlfluff{}~\cite{SQLFluff} as baselines. \sqlcheck{} originally combined a non-validating parser with regular expression matching to detect anti-patterns~\cite{dintyalaSQLCheckAutomatedDetection2020}. However, its current open-source implementation relies only on regular expressions. \sqlfluff{} uses a multi-dialect parser to parse input queries into ASTs, which are then analyzed for anti-patterns. While \sqlcheck{} avoids parsing failures by using text-based analysis instead of ASTs, this approach leads to potentially lower accuracy. In contrast, \sqlfluff{} has higher accuracy, but requires significant manual effort to support new dialects. Since \sqlfluff{} allows users to configure the SQL dialect, we report results for its ANSI mode (\emph{SF-ANSI}) and TSQL mode (\emph{SF-TSQL}). Lastly, we implemented a purely LLM-based baseline using GPT-4.1 to investigate whether an end-to-end LLM-based approach is sufficient for this task. We provided the LLM natural language rule descriptions and prompted it to output anti-patterns.

{

\begin{table}

    \centering    \caption{List of Selected Linter Rules}
    \footnotesize
    \vspace{-8pt}

    \begin{tabular}{@{}lll@{}}
    \toprule
    \textbf{Source} & \textbf{ID} & \textbf{Description}   \\ 
    \midrule
    \multirow{11}{*}{{\sqlfluff{}}}
        & AM01 & Ambiguous use of \texttt{DISTINCT} with \texttt{GROUP BY}. \\
        & AM06 & Inconsistent column references in \texttt{GROUP/ORDER BY} clauses.\\
        & AL02 & Implicit aliasing of columns. \\
        & AL03 & Column expression without alias. \\
        & AL04 & Table aliases should be unique within each clause. \\
        & AL05 & Tables should not be aliased if that alias is not used. \\ 
        & AL08 & Column aliases should be unique within each clause. \\
        & AL09 & Column aliases should not alias to themselves. \\
        & CV04 & Use consistent syntax to express ``count number of rows''.  \\
        & CV05 & Comparisons with \texttt{NULL} should use \texttt{IS} or \texttt{IS NOT}. \\
        & RF01 & Referencing objects not present in the \texttt{FROM} clause. \\
    \midrule
    \multirow{3}{*}{{Both tools}}
        & AM02 & \texttt{UNION [DISTINCT|ALL]} is preferred over \texttt{UNION}. \\
        & AM04 & Query produces an unknown number of result columns. \\
        & AM08 & Implicit cross join detected.  \\
    \bottomrule
    \end{tabular}
    \label{tb:linter-rules}
\end{table}
}
\paragraph{Rule selection}

We systematically selected a set of fourteen linter rules for the evaluation, presented in Table~\ref{tb:linter-rules}. We first selected all three rules that are supported by both \sqlfluff{} and \sqlcheck{}. For a more comprehensive evaluation, we incorporated eleven additional rules from \sqlfluff{}. We chose these rules from \sqlfluff{}'s \emph{core} rule set, which are defined by the developers as \emph{``Stable, applies to most dialects, could detect a syntax issue, and is not too opinionated toward one style''}~\cite{SQLFluff}. Within the core rule set, we excluded those addressing layout issues (\eg{} incorrect indentation), symbol inconsistencies (\eg{} extra commas), and capitalization inconsistencies (\eg{} keyword capitalization), as \sqlflex{} does not preserve this information during parsing. For brevity, we will refer to each rule by its ID in Table~\ref{tb:linter-rules} in subsequent paragraphs.


\paragraph{Dataset}
{We selected the SESD dataset~\cite{hazoom2021TexttoSQLWildNaturallyOccurring}, which consists of human-authored SQL queries with diverse TSQL-specific features collected from \emph{Stack Exchange}, and we expected some of them to include anti-patterns targeted by SQL linters.} We removed duplicate queries and excluded queries that failed to parse using a TSQL parser. {Finally, we obtained 1,916 TSQL queries, including 111 queries with window functions and 1,030 queries with joins.}


Since \sqlfluff{} already supports the TSQL dialect, we constructed reliable ground-truth annotations (\ie{} the list of potential anti-patterns for each query) by initially running both \sqlflex{} and \sqlfluff{} on the entire dataset. Cases where both tools detected the same anti-patterns were considered the ground-truth. For discrepancies between the two tools, we manually reviewed and confirmed each instance to ensure accurate labeling. {To further validate the constructed ground truth, we randomly sampled 50 labeled queries and manually annotated them. We observed no discrepancies between these annotations and the constructed labels. Based on this sample, a Clopper–Pearson confidence interval shows that, with 95\% confidence, the true error rate is below approximately 6\%.}

\paragraph{Metrics}
We use \emph{Precision}, \emph{Recall}, and \emph{F1 score} as evaluation metrics~\cite{Russell2020AI}. The metrics are computed at the anti-pattern level. For example, if the ground truth for a query is \{AM01, AL02, RF01\}, but \sqlflex{} reports \{AM01, AL03\}, we count one true positive (AM01), one false positive (AL03), and one false negative (RF01). Here, true positives refer to correctly detected anti-patterns, false positives to incorrectly reported ones, and false negatives to missed anti-patterns. For all three metrics, higher values indicate better performance. Since \sqlcheck{} supports only three of the selected rules, we compute its metrics based on those three rules.

\begin{figure}
    \centering
    \includegraphics[width=0.7\linewidth]{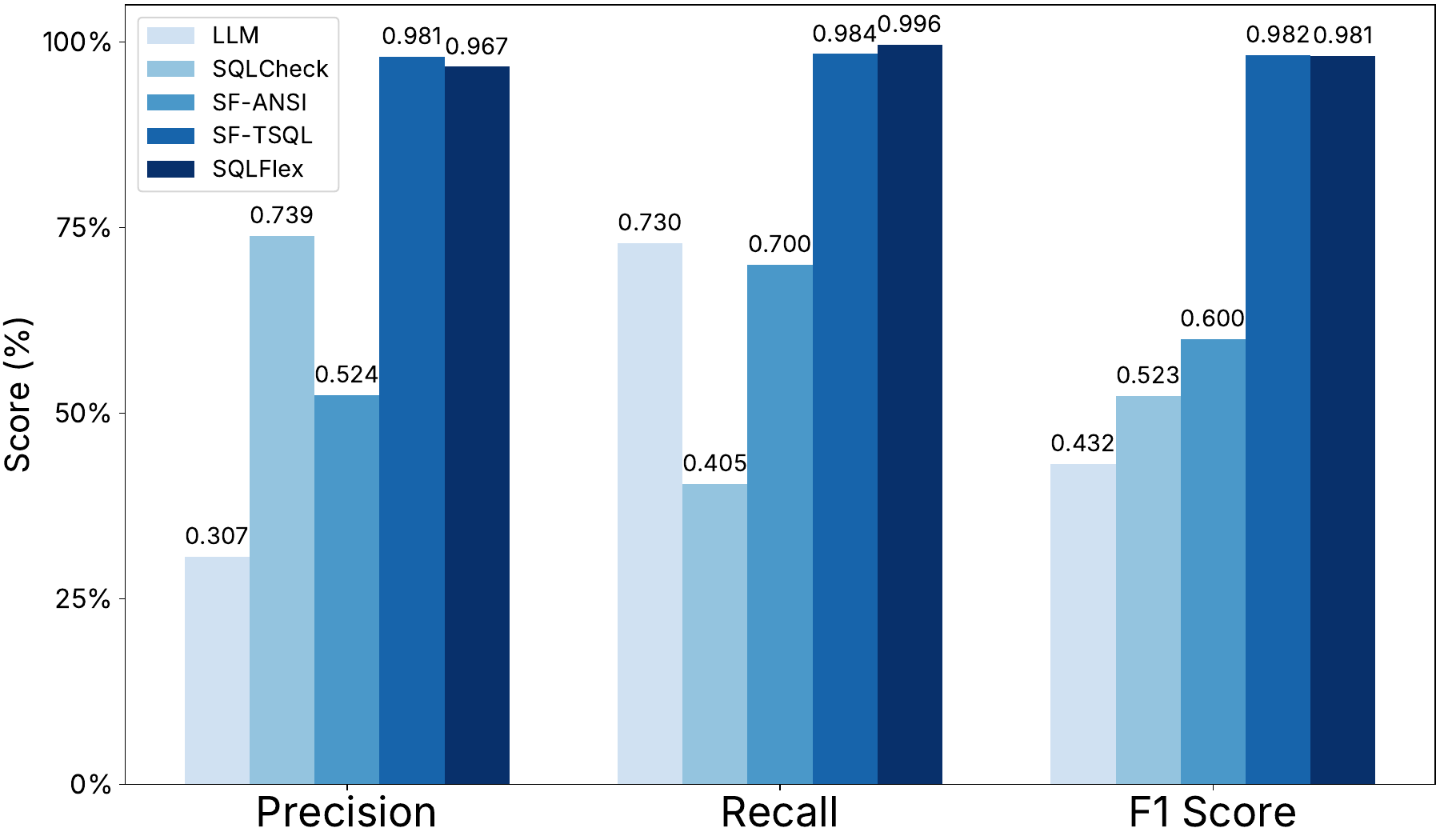}
    \vspace{-8pt}
    \caption{Overall performance comparison of linters}
    \label{fig:linter-acc}
\end{figure}

\paragraph{Results}
Figure~\ref{fig:linter-acc} shows the overall results for all selected linter rules. \sqlflex{} significantly outperforms the purely LLM-based approach, \sqlcheck{}, and \emph{SF-ANSI} across all three metrics. The purely LLM-based linter suffers from high false positive rates, likely due to the ambiguity of natural language rule descriptions and the lack of effective validation mechanisms, which highlights the importance of AST-based analysis. Similarly, \sqlcheck{}, which relies on text-based analysis without accounting for SQL’s structural semantics (\eg{} column names), also performs poorly. \emph{SF-ANSI} frequently failed to parse queries in the SESD dataset due to TSQL-specific syntax, such as using square brackets to enclose identifiers with special characters or reserved keywords (\eg{} \codeIn{SELECT t AS [SELECT]}), leading to low linting accuracy. In contrast, \sqlflex{} was able to handle such syntax correctly via the hybrid segmentation approach. 

When compared with \emph{SF-TSQL}, \sqlflex{} achieves slightly lower precision (96.69\% vs. 98.06\%), higher recall (99.63\% vs. 98.42\%), and comparable F1 score (98.14\% vs. 98.24\%). The errors made by \sqlflex{} are mainly due to imprecise segmentation by the LLM when faced with ambiguous syntax in queries. For example, in ``\codeIn{SELECT Count FROM t} \ldots'', the LLM misinterpreted \codeIn{Count} as an expression (resembling function \codeIn{COUNT()}) rather than a column name, causing the linter to incorrectly report the anti-pattern ``AL03''. \emph{SF-TSQL} also had occasional parsing failures that led to incorrect results. For example, it failed to parse the query ``\codeIn{SELECT Class AS Class} \ldots'', causing it to miss the anti-pattern ``AL09''. Overall, the linter implemented using \sqlflex{} matched the performance of \sqlfluff{}, with significantly less manual effort.




\subsection{Test-Case Reduction}\label{sec:usecase-reduction}
Test-case reduction is a query rewriting application. Automated DBMS testing tools, such as \sqlancer{}~\cite{SQLancer} and SQLsmith~\cite{sqlsmith}, often generate long and complex queries to expose bugs. However, these queries are often too complicated for developers to identify the underlying issues. Consider the NoREC oracle~\cite{Rigger2020NoREC}, which identifies logic bugs by comparing results between two semantically equivalent queries that differ only in whether query optimization is applied. The resulting bug reports typically contain complex expressions~\cite{Rigger2020NoREC, zhong2025scalingautomateddatabasetesting, yu2022sqlright} that make it difficult for developers to fix potential bugs. Therefore, reducing test cases while preserving bug-triggering behavior is essential for efficient bug-fixing~\cite{sun2018perses, shi2015comparingAndCombining, pan2023ATMMinimization}.

\paragraph{Baselines} 
\sqless{}~\cite{lin2024SQLessDialectAgnosticSQL} is a state-of-the-art test case reducer designed specifically for dialect-agnostic SQL reduction. It uses AST-based strategies, such as removing clauses and simplifying expressions, to iteratively reduce queries. After each reduction, \sqless{} executes the reduced query with the test oracle to verify whether the bug can still be triggered. When encountering a query containing dialect-specific features, \sqless{} uses {ANTLR's syntax-based error recovery algorithm} to generate new grammar rules. With these added grammar rules, \sqless{} can parse previously unsupported queries. We compare a test-case reduction tool implemented in \sqlflex{} with \sqless{} as the baseline.

\paragraph{Dataset}
\sloppy{}
We constructed a new dataset of bug-triggering queries in SQLite and MySQL to evaluate \sqlflex{}. A possible alternative would have been SQLess's dataset. However, it is not publicly available and is imbalanced, containing over 32,000 MySQL-compatible cases, but only 143 SQLite cases. Thus, we used \sqlancer{}, a widely used DBMS testing tool, to generate bug-triggering queries in prior versions of SQLite (v3.28.0) and MySQL (v8.2.0). We increased the test case complexity by setting \sqlancer{}'s expression-depth parameter to 5, then collected the first 1,500 bug-inducing test cases for each dialect, retaining only those found by the NoREC oracle. {The final dataset includes 1,166 SQLite and 1,491 MySQL test cases, among which 583  contain subqueries, and 767 have joins.} Notably, \sqless{} already supports MySQL as it uses the MySQL grammar, but needs its adaptive parser to handle SQLite-specific features. 

\paragraph{Implementation}
We adopted the test-case reduction algorithm described in \sqless{}~\cite{lin2024SQLessDialectAgnosticSQL}. Specifically, we implemented expression reduction strategies (\eg{} removing the left operand of a logical operator) and clause reduction strategies (\eg{} removing the \codeIn{ORDER BY} clause) on the ASTs generated by \sqlflex{}. We implemented only the strategies relevant to our dataset and sufficient for effective simplification. Since \sqless{} applies strategies enumeratively, this selective implementation does not bias the results. For a fair comparison, we disabled additional strategies in \sqless{}.


\paragraph{Metrics}
Following \sqless{}, we used \emph{Average Simplification Ratio} (AvgSimRatio) and \emph{Maximum Simplification Ratio} (MaxSimRatio) as evaluation metrics~\cite{lin2024SQLessDialectAgnosticSQL}. These metrics measure the percentage reduction in the number of tokens between the original and reduced queries, with higher values indicating a more effective reduction. 


\paragraph{Results}
Table~\ref{tb:reduction-ratios} compares the simplification ratios achieved by \sqless{} and \sqlflex{}. Overall, \sqlflex{} performs better than \sqless{}. For the MySQL test cases, both tools achieved comparable performance, showing that \sqlflex{}, despite using a minimal grammar---with 61.82\% of reduced queries requiring segmentation---can match the effectiveness of a full-grammar-based tool. For the SQLite test cases, \sqlflex{} significantly outperforms \sqless{}. This is primarily because \sqless{} struggled with SQLite-specific features, while \sqlflex{} effectively handled them. The adaptive parser in \sqless{} appends all unrecognized tokens as terminal symbols without capturing their semantic meaning. Moreover, for complex queries, the adaptive parser often generates incorrect grammar rules, resulting in ineffective reductions, as the reduction rules fail to interpret the tree nodes. In contrast, although 92.78\% of reduced queries were not supported by \sqlflex{}'s grammar, \sqlflex{} still generates ASTs that preserve the semantic structure of the SQL query. Its anchor-based expression segmentation approach further enables robust handling of complex expressions, leading to more effective reductions.



\begin{table}
  \centering
  \small
  \caption{Simplification ratios by DBMS}
      \vspace{-8pt}

  \begin{tabular}{@{}lrrrr@{}}
    \toprule
    \multirow{2}{*}{\textbf{DBMS}}
      & \multicolumn{2}{c}{\textbf{MaxSimRatio}}
      & \multicolumn{2}{c}{\textbf{AvgSimRatio}} \\
    \cmidrule(lr){2-3} \cmidrule(lr){4-5}
      & \textbf{\sqless{}} & \textbf{\sqlflex{}}
      & \textbf{\sqless{}} & \textbf{\sqlflex{}} \\
    \midrule
    MySQL     & \textbf{87.92\%} & \textbf{87.92}\% & 6.27\% & \textbf{7.53\%} \\
    SQLite   & 86.83\% & \textbf{94.75}\% & 1.83\% & \textbf{18.91\%} \\
    \bottomrule
  \end{tabular}
  \label{tb:reduction-ratios}
\end{table}

\section{Standalone Evaluation}

In this section, we further investigate \sqlflex{} itself, that is, whether \sqlflex{} can effectively parse queries. Specifically, we aimed to answer the following questions:

\begin{itemize}[leftmargin=*]
    \item \textbf{Q1.} How effectively does \sqlflex{} handle dialect-agnostic parsing of SQL queries (see Section~\ref{sec:eval-ast})?
    \item \textbf{Q2.} How do \sqlflex{}'s key components contribute to its overall effectiveness (see Section~\ref{sec:ablation})?
    \item \textbf{Q3.} How efficiently does \sqlflex{} handle SQL queries with dialect-specific features (see Section~\ref{sec:eval-efficiency})?
    \item \textbf{Q4.} Can \sqlflex{} be extended to parse Data Definition Language (DDL) statements, such as \codeIn{CREATE TABLE} (see Section~\ref{sec:ddl})?

\end{itemize}



\paragraph{Dataset}
To investigate whether \sqlflex{} can handle diverse and niche dialect-specific features, we built a large-scale benchmark dataset extending the dialect types covered in Section~\ref{sec:use-case}. {To the best of our knowledge, there is no publicly available benchmark suitable for the evaluation of dialect-agnostic SQL parsing. While there are existing query benchmarks such as TPC-H and SQLStorm~\cite{schmidt25sqlstorm}, these benchmarks are designed for DBMS performance evaluation and primarily contain standard SQL features.} Our benchmark consists of queries of eight different dialects, which we collected from the corresponding DBMS’s test suite. These test suites typically include statements that reflect the unique features of their respective systems~\cite{zhong2024UnderstandingReusingTest}, making them more suitable for the evaluation of \sqlflex{}'s dialect-agnostic parsing ability.

We selected the SQL dialects and the DBMSs that implement them based on several factors: (1) availability of open-source test suites; (2) diversity in usage and functionality, including a combination of well-established systems (\eg{} PostgreSQL and MySQL) and emerging or specialized ones (\eg{} DuckDB and SurrealDB)~\cite{db-ranking}; (3) distinct dialect-specific features, avoiding largely compatible systems like MySQL and MariaDB that share a common codebase. {Table~\ref{tb:dataset} summarizes the datasets, including the number of queries and the occurrence of complex query features (\ie{} window functions, CTEs, subqueries, and joins).} The datasets cover a broad range of SQL features, evaluating \sqlflex{} comprehensively.

We built the benchmark dataset by extracting \codeIn{SELECT} statements from test suites. To ensure syntactic validity, we excluded queries that failed to parse using dialect-specific parsers. Since many of the extracted queries had similar structures or features, we applied a keyword-based deduplication heuristic to retain a diverse subset. SQL parsing depends primarily on keywords~\cite{sqlHasProblems24Shute}, making keyword diversity a useful proxy for syntactic variety. We treated queries with identical keyword sets as duplicates, retaining only the longest one, assuming it to be the most complex. For H2, since we were unable to obtain a ready-to-use H2-specific parser, we used JSQLParser, which supports H2 syntax. For SurrealDB, as we were unable to obtain its keyword list, we kept all queries in its dataset.

\begin{table}[]
    \centering
    \caption{List of datasets and query features used in the standalone evaluation (WF: Window Function, CTE: Common Table Expression, SubQ: Subquery)}
    \vspace{-8pt}

    \footnotesize
    
    {
    \begin{tabular}{@{}lrrrrr@{}}
    \toprule
    \textbf{Dataset} &
    \textbf{Queries} &
    \textbf{WFs} &
    \textbf{CTEs} &
    \textbf{SubQs} &
    \textbf{Joins} \\
    \midrule
    DuckDB Test Suite
        & 2,092 & 185 & 22 & 496 & 284 \\
    PostgreSQL Test Suite
        & 1,207 & 139 & 0 & 162 & 128 \\
    MySQL Test Suite
        & 1,039 & 3 & 0 & 67 & 77 \\
    ClickHouse Test Suite
        & 2,485 & 69 & 1 & 573 & 513 \\
    H2 Test Suite
        & 358 & 21 & 0 & 67 & 50 \\
    CockroachDB Test Suite
        & 1,369 & 100 & 2 & 287 & 202 \\
    Cassandra Test Suite
        & 150 & 0 & 0 & 0 & 0 \\
    SurrealDB Test Suite
        & 351 & 0 & 0 & 5 & 0 \\
    \midrule
    \textbf{Total}
        & \textbf{9,051} & \textbf{517} & \textbf{25} & \textbf{1,657} & \textbf{1,254} \\
    \bottomrule
    \end{tabular}
    }
    \label{tb:dataset}
\end{table}

\subsection{Q1: Effectivness and Reliability}\label{sec:eval-ast}

In this section, we show \sqlflex{}'s effectiveness and reliability in parsing queries in diverse SQL dialects.

\paragraph{Baselines}
We compared \sqlflex{} against representative baselines. To the best of our knowledge, no existing academic work has proposed a general-purpose, dialect-agnostic SQL parser. The most closely related work is \sqless{}~\cite{lin2024SQLessDialectAgnosticSQL}. As SQLess is applicable only to test-case reduction, we evaluate it on this use case in Section~\ref{sec:usecase-reduction}. Thus, we compare \sqlflex{} against two open-source tools, which are representatives of dialect-specific parsers and multi-dialect parsers, and a purely LLM-based baseline parser:

\begin{itemize}[leftmargin=*]
   \item \textbf{PostgreSQL Parser:} We chose the PostgreSQL parser as the dialect-specific parser baseline. We used \pglast{}~{(v6.10)}~\cite{pglast}, a Python package of the official PostgreSQL parser. While dialect-specific parsers exist for other dialects, we chose to focus on PostgreSQL for two reasons. First, it is a widely used dialect with strong conformance to SQL standards. Second, we expect similar results for other dialect-specific parsers, as they share the same limitation. Specifically, we expect them to perform perfectly on their target dialect, but fail to parse specific features of other dialects. \looseness=-1
    \item \textbf{\sqlglot{} {v25.22.0}:} \sqlglot{}~\cite{SQLGlot} is a state-of-the-art multi-dialect parser supporting over 30 dialects. {We evaluated two configurations of \sqlglot{}, which are its standard mode (\textbf{\sqlglot{}$_s$}), which parses queries using a superset of all supported dialects, and its dialect-specific mode, where the dialect is supported (\textbf{\sqlglot{}$_d$}). \sqlglot{} is expected to perform well on its explicitly supported dialects due to its extensive manual implementation effort, and better than \pglast{} on unsupported dialects in its standard mode. In other words, we set SQLGlot’s configuration so that it is expected to perform best for each of the dialects.}

   \item \textbf{LLM-based Baseline Parser:} We implemented a simple baseline that uses an LLM (GPT-4.1 API) to generate an AST in JSON format directly. We provide the same AST definition as used by \sqlflex{}. We use a rule-based printer to convert the AST into a query to prevent hallucinations from an LLM-based pretty-printer and align with \sqlflex{} also using a rule-based printer.
\end{itemize}

\paragraph{Metrics}

Evaluating the semantic correctness of ASTs in a multi-dialect setting is inherently challenging for two reasons. First, different parsers adopt different AST representations, which can lead to structurally different, yet semantically equivalent ASTs for the same query. For example, \pglast{} represents both the \codeIn{CAST} function and the \codeIn{::} operator using a unified type-casting node. Similarly, \sqlglot{} represents both \codeIn{IF} and \codeIn{CASE} expressions using a single expression type. Second, constructing reliable ground-truth ASTs is challenging due to the inaccessibility of official reference parsers; specifically, while some DBMSs offer official standalone parser packages, for example \pglast{}, which is derived directly from PostgreSQL’s source code, many others only embed their parsers within the database engine (\eg{} DuckDB). Extracting such embedded parsers would incur significant engineering effort.

To address the challenge in evaluating SQL parsers, we adopt property-based testing~\cite{quickcheck2000}, a technique that checks whether a system behaves as expected according to specific properties. In particular, we used a widely used round-trip property to measure parser correctness~\cite{danielsson2013correctPrettyPrinting, geest2017packetParsing, rendel2010unifyingParsingPrettyPrint}, and satisfying such a property is useful for developing query rewriting tools~\cite{sqlparser-rs}. The round-trip property checks whether a query can be parsed into an AST and printed back to the same query as the input, offering a practical way to evaluate parsers without needing ground-truth labels. 

We introduce \emph{Query Round-trip (Q-RT) Rate}, the metric for our evaluation. Q-RT is an extension of the round-trip property, which mitigates inaccurate evaluation results caused by normalization applied by baselines. Given an initial query $Q_0$, we define intermediate parsing and printing steps as follows:
\begin{align}
    Q_1 = \text{Print}(\text{Parse}(Q_0)), \quad 
    Q_2 = \text{Print}(\text{Parse}(Q_1))
\end{align}
Based on these steps, we define the Q-RT property, which checks that the queries remain identical through each parse-print cycle (\ie{} \(Q_0 = Q_1 = Q_2\)). It provides insights on whether \sqlflex{} can consistently parse a query into an AST and transform the AST back to an executable query. We relax Q-RT for \pglast{} and \sqlglot{} to only require \(Q_1 = Q_2\), to avoid disadvantaging them due to normalization (\eg{} \pglast{} converts \codeIn{::} operators into \codeIn{CAST} functions). Additionally, we ignore whitespace and parentheses during comparison to avoid false alarms from formatting differences. \emph{Q-RT Rate} is calculated as the ratio of the number of queries satisfying Q-RT to the total number of queries in each dataset. Higher values indicate better dialect-agnostic parsing effectiveness and reliability. Due to space limits, we present additional results and analysis of AST-level consistency using an AST round-trip property in the appendix. Note that while the round-trip property cannot reliably detect incorrect ASTs, it serves as a necessary condition for practical parser correctness. We excluded \sqless{} from this comparison, as its implementation guarantees round-trip equivalence, but produces ASTs with limited practical utility, as user-defined rules often fail to interpret the AST. The selected baselines satisfy the round-trip property while also producing usable ASTs, making them suitable for this evaluation. 


In addition to Q-RT, we further evaluate the parsing correctness of \sqlflex{} by assessing semantic correctness, where \pglast{} and the PostgreSQL queries naturally serve as the ground-truth. Following a methodology of previous work on evaluating correctness of DDL statement parsing~\cite{till24schemapile}, we extract a core set of semantic elements from a ground-truth AST and compare whether these elements match those extracted from the ASTs produced by \sqlflex{} and \sqlglot{}. Specifically, at the clause level, we extract column names from the \codeIn{SELECT} clause, table names from the \codeIn{FROM} clause, aliases and their targets, as well as joined table names and join types from the \codeIn{JOIN} clauses. At the expression level, we extract operators, their literal operands, and their precedence. Since different parsers may normalize certain operators differently, we restrict extraction to a predefined set of common operators to avoid counting semantically equivalent forms as mismatches. For each semantic element, we compute the percentage of ASTs whose extracted results overlap with the ground truth. We refer to this complementary metric as the \emph{Semantic Match (SM) Rate}. Note that the SM rate also remains a best-effort correctness metric, but it provides a correctness reference, complementing the self-referencing round-trip metric.

\paragraph{Dialect-agnostic effectiveness (Q1)}
Table~\ref{tb:accuracy} presents the results comparing \sqlflex{} against baselines in Q-RT rate, and reports the geometric mean~\cite{fleming1986GeometricMean} in the last row. For each dialect, the \sqlflex{} results are averaged over three runs, with standard deviations below 0.6\%, indicating consistent performance. Overall, the results show that \sqlflex{} outperforms all baselines. \pglast{} achieves an expected perfect score for the PostgreSQL dataset, but across all dialects, it only has a geometric mean of 65.94\% Q-RT. The standard mode of \sqlglot{} (\sqlglot{}$_s$) performs slightly better, with a geometric mean of 74.43\%. The dialect-specific mode (\sqlglot{}$_d$) achieves a higher geometric mean of 93.26\%, but only on dialects it explicitly supports. This improved performance is due to its manually implemented parsing rules for those dialects. However, it requires substantial human effort and still fails to capture some unique features. The LLM-based baseline parser, although requiring no manual effort, performs poorly at generating consistent ASTs (46.48\% geometric mean Q-RT). In contrast, \sqlflex{} outperforms all baselines with a 96.37\% geometric mean Q-RT, and ranges from 91.55\% to 100\% Q-RT across all eight dialects.

We inspected all failed cases. We find that for \pglast{} and \sqlglot{}, almost all failures are due to parsing errors caused by dialect-specific features. For the LLM-based baseline, failures occur due to inconsistent end-to-end generation that produces different results across the AST generation and printing stages. For \sqlflex{}, the few Q-RT failures are mostly caused by limitations in handling dialect-specific features in the rule-based pretty-printer. A typical example comes from CockroachDB, which allows explicit index annotations like \codeIn{``table@idx''}~\cite{cockroachdb-select-index} in table references. In certain queries in the dataset, this is written as \codeIn{``cb@w''}. Since the generic identifiers (\codeIn{cb} and \codeIn{w}) provide no semantic information, the segmenter may incorrectly interpret \codeIn{cb@w} as a schema-qualified table name. After pretty-printing, the output becomes \codeIn{cb.w}, which fails to match the original input. Similarly, in the SurrealDB dataset, clauses like \codeIn{``GROUP ALL''} are incorrectly pretty-printed as \codeIn{``GROUP BY ALL''}. Lastly, a small portion of AST produced by \sqlflex{} (0--2.5\%) contained \codeIn{Unsegmented} nodes, which we counted as failed cases. These typically arise from LLM misinterpretations in complex queries (\eg{} multiple subqueries) and fail to be repaired.

\paragraph{Semantic match rates (Q1)}
We report the SM rates on PostgreSQL queries in Table~\ref{tb:against-pglast}. Overall, \sqlflex{} achieves the highest SM rates across all evaluated semantic elements, exceeding 94\% at both the clause level and the expression level, and outperforming both variants of \sqlglot{}. While the overall trend of SM is consistent with Q-RT, SM additionally captures certain semantic mismatches that Q-RT fails to detect. For example, \sqlflex{} incorrectly interpreted the expression \codeIn{int2 `2'} in the \codeIn{SELECT} clause as a column reference, whereas it was actually an implicit type-conversion expression that casts \codeIn{`2'} to type \codeIn{int2}. This mismatch was missed by Q-RT, because the printed columns were identical despite the differing semantics. Although such cases can be challenging even for LLM-based parsing, their rarity is reflected in the high SM rates, indicating that \sqlflex{} remains effective in practice.

\begin{table}[]
    \centering
    \footnotesize
    \caption{Query Round-Trip Rates. Bold indicates the best score for each dataset (excluding \pglast{} on PostgreSQL)}
        \vspace{-4pt}

    \begin{tabular}{@{}lrrrrr@{}}
    \toprule
    \textbf{Dataset} &
      \textbf{LLM} &
      \textbf{\pglast{}} &
      \textbf{\sqlglot{}$_s$} &
      \textbf{\sqlglot{}$_d$} &
      \textbf{\sqlflex{}} \\ 
    \midrule
         DuckDB       & 51.53\%  & 72.42\%  & 92.97\%  & 93.88\%  & \textbf{95.86\%} \\
         PostgreSQL   & 49.30\%  & 100.00\% & 86.67\%  & 87.99\%  & \textbf{96.88\%} \\
         MySQL        & 66.41\%  & 76.42\%  & 86.91\%  & 97.40\%  & \textbf{99.39\%} \\
         ClickHouse   & 32.27\%  & 47.00\%  & 68.61\%  & 94.00\%  & \textbf{94.67\%} \\
         H2           & 56.15\%  & 81.01\%  & 94.69\%  & –        & \textbf{97.86\%} \\
         CockroachDB  & 43.46\%  & 69.98\%  & 70.93\%  & –        & \textbf{95.06\%} \\
         Cassandra    & 60.00\%  & 74.00\%  & 76.00\%  & –        & \textbf{100.00\%} \\
         SurrealDB    & 27.35\%  & 32.76\%  & 38.46\%  & –        & \textbf{91.55\%} \\
    \midrule
         Mean      & 46.48\%  & 65.94\%  & 74.43\%  & 93.26\%  & \textbf{96.37\%} \\
    \bottomrule
    \end{tabular}
    \label{tb:accuracy}
\end{table}

\begin{table}[]
    \centering
    \footnotesize
    \setlength{\tabcolsep}{4pt}
    \caption{SM rates on the PostgreSQL dataset. Bold indicates the best. (Opt.: Operator, Opn.: Operand, Prec.: Precedence)}
    \vspace{-4pt}
    {
    \begin{tabular}{@{}lrrrrrrr@{}}
    \toprule
    \textbf{Parser} & \textbf{Tables} & \textbf{Cols} & \textbf{Alias} & \textbf{Joins} & \textbf{Opt.} & \textbf{Opn.} & \textbf{Prec.} \\
    \midrule
    \sqlflex{}       
      & \textbf{94.86\%} 
      & \textbf{98.09\%} 
      & \textbf{95.86\%} 
      & \textbf{97.93\%} 
      & \textbf{97.68\%} 
      & \textbf{95.44\%} 
      & \textbf{94.20\%} \\
    \sqlglot{}$_d$ 
      & 79.37\% 
      & 88.40\% 
      & 85.17\% 
      & 86.33\% 
      & 86.58\% 
      & 84.18\% 
      & 84.34\% \\
    \sqlglot{}$_s$ 
      & 77.46\% 
      & 86.58\% 
      & 83.26\% 
      & 84.51\% 
      & 84.76\% 
      & 82.35\% 
      & 82.52\% \\
    
    \bottomrule
    \end{tabular}
    }
    \label{tb:against-pglast}
\end{table}

\subsection{Q2: Ablation Study}\label{sec:ablation}
In this section, we investigate the contributions of important components of \sqlflex{}.


\paragraph{Methodology}

We compared four variants of \sqlflex{} to assess the impact of its key components:
\begin{itemize}[leftmargin=*]
    \item \textbf{SQL-92:} We removed the entire LLM-based segmenter to demonstrate the effectiveness of segmentation overall.
    \item \textbf{\sqlflex{}$_v$:} We removed the validation and repair component to examine its contribution to parsing correctness.
    \item \textbf{\sqlflex{}$_a$:} We removed anchor-based expression segmentation, letting the segmenter process an entire expression.
    \item \textbf{\sqlflex{}$_m$:} We replaced GPT-4.1 with GPT-4.1-mini as the base model to evaluate whether \sqlflex{} remains effective with a more lightweight LLM.
\end{itemize}
We evaluated all variants on eight datasets using the Q-RT rate and also report the number of LLM calls for expression segmentation in \sqlflex{} and \sqlflex{}$_a$.


\paragraph{Component contributions (Q2)}
We compare the geometric mean across the eight datasets with results shown in Table~\ref{tb:ablation-full}. Removing the entire LLM-based segmenter (SQL-92) results in a drastic performance drop to 38.39\% Q-RT, highlighting that LLM-based segmentation is highly effective when the grammar-based parser fails to parse dialect-specific features. Removing only the validation and repair module (\sqlflex{}$_{v}$) leads to a 7.60\% Q-RT drop, showing that while the core segmentation is already reliable, validation further improves robustness by correcting edge cases. Removing only the anchor-based segmentation (\sqlflex{}$_{a}$) causes a 5.83\% Q-RT drop, indicating that our approach improves expression handling effectiveness. Moreover, \sqlflex{}$_{a}$ uses 36.61\% more LLM calls in total, showing that the anchor-based strategy improves efficiency by avoiding redundant segmentation of known operators. In queries where the expressions are more complex, we believe the anchor-based strategy would be even more effective and efficient in parsing those expressions. Finally, \sqlflex{}$_m$, which replaces GPT-4.1 with GPT-4.1-mini, achieves a mean Q-RT of 92.58\%, only 3.79\% below \sqlflex{}, showing that \sqlflex{} remains accurate even for lightweight models.

\begin{table}[]
    \centering
    \footnotesize
    \caption{Ablation study results in Q-RT rates. Bold indicates the best score for each dataset.}
        \vspace{-4pt}

    \begin{tabular}{@{}lrrrrr@{}}
    \toprule
    \textbf{Dataset} & \textbf{SQL-92} & \textbf{\sqlflex{}$_v$} & \textbf{\sqlflex{}$_a$} & \textbf{\sqlflex{}$_m$} & \textbf{\sqlflex{}} \\
    \midrule
    DuckDB       & 31.98\% & 88.91\% & 86.42\% & 90.68\% & \textbf{95.86\%} \\
    PostgreSQL   & 36.54\% & 90.89\% & 85.25\% & 95.44\% & \textbf{96.88\%} \\
    MySQL        & 62.57\% & 95.00\% & 93.65\% & 96.92\% & \textbf{99.39\%} \\
    ClickHouse   & 37.23\% & 90.38\% & 90.70\% & 92.31\% & \textbf{94.67\%} \\
    H2           & 54.19\% & 93.30\% & 91.62\% & 96.09\% & \textbf{97.86\%} \\
    CockroachDB  & 34.34\% & 72.61\% & 86.71\% & 85.17\% & \textbf{95.06\%} \\
    Cassandra    & 53.34\% & 100.00\% & 100.00\% & 98.67\% & \textbf{100.00\%} \\
    SurrealDB    & 17.47\% & 82.05\% & 90.88\% & 86.34\% & \textbf{91.55\%} \\
    \midrule
    Mean         & 38.39\% & 88.77\% & 90.55\% & 92.58\% & \textbf{96.37\%} \\
    \bottomrule
    \end{tabular}
    \label{tb:ablation-full}
\end{table}

\subsection{Q3: Efficiency}\label{sec:eval-efficiency}

In this section, we investigate \sqlflex{}'s efficiency in query parsing and the overhead of LLM calls.
\paragraph{Methodology}
We assessed the performance of \sqlflex{} on each dataset by measuring parsing time and the number of LLM calls (inclusive of repair attempts). Since the base grammar can affect the average efficiency (\ie{} more grammar rule matches indicate fewer LLM calls), we use two grammar configurations to provide a more comprehensive view on the efficiency of \sqlflex{}:
\begin{itemize}[leftmargin=*]
\item \textbf{SQL-92:} The standard implementation of \sqlflex{}.
\item \textbf{SQL:2003} A newer SQL standard to SQL-92, which extended SQL-92 by adding more features such as window functions.

\end{itemize}

For each configuration, we report the average parsing time ($T_{avg}$) and total number of LLM calls ($N_{LLM}$) per dialect. We also measure the average parsing time specifically for queries that required segmentation ($T_{LLM}$).

\paragraph{Efficiency (Q3)}

We present the results in Table~\ref{tb:efficiency-metrics}. Under the standard SQL-92 configuration, each query takes on average 3.67 seconds to parse, while queries requiring segmentation take 6.39 seconds. Given the limited coverage of the base grammar and the prevalence of dialect-specific features in the collected queries, frequent LLM invocations are expected (as shown in Section~\ref{sec:ablation}), leading to increased parsing overhead. After extending support to SQL:2003 features, we observe a 21.6\% reduction in the number of LLM calls and a 16.62\% decrease in average parsing time. This improvement can be attributed to the inclusion of additional grammar rules that capture commonly used SQL features. Moreover, with more anchors introduced in SQL:2003, such as in window functions, segmentation itself achieves a further 7.04\% speed-up. {Under the SQL:2003 configuration, queries that required segmentation incurred a median of 1 to 6 LLM calls across the eight dialects. The worst case that we encountered was an artificially constructed PostgreSQL query containing 90 type-cast operators \codeIn{::} to an unknown \codeIn{inet} type, which triggered 94 LLM calls. Note that this worst case is an outlier rather than typical behavior, and the number of LLM calls is highly workload-specific. In practice, users can extend the grammar with frequently used SQL features, such as the \codeIn{::} operator to further reduce parsing overhead, while still having the flexibility to handle queries beyond the defined grammar.}

\begin{table}[]
    \centering
    \footnotesize
    \caption{Efficiency metrics for query parsing with \sqlflex{}. Time is measured in seconds.}
        \vspace{-4pt}

    \label{tb:efficiency-metrics}
    \begin{tabular}{@{}lrrrrrrrr@{}}
        \toprule
        \multirow{2}{*}{\textbf{Dataset}} 
            & \multicolumn{3}{c}{\textbf{SQL-92}} 
            & \multicolumn{3}{c}{\textbf{SQL:2003}} \\
        \cmidrule(lr){2-4} \cmidrule(lr){5-7}
            & \textbf{$T_{avg}$} & \textbf{$T_{LLM}$} & \textbf{$N_{LLM}$}
            & \textbf{$T_{avg}$} & \textbf{$T_{LLM}$} & \textbf{$N_{LLM}$} \\
        \midrule
        DuckDB      & 6.09 & 8.96 & 10101 & 5.26 & 9.08 & 7922 \\
        PostgreSQL  & 6.81 & 10.74 & 5422 & 5.55 & 10.20 & 4276 \\
        MySQL       & 1.94 & 5.13 & 1930 & 1.87 & 5.52 & 1721 \\
        ClickHouse  & 4.38 & 7.02 & 6912 & 3.76 & 6.32 & 6148 \\
        H2          & 2.93 & 6.35 & 795 & 1.97 & 5.31 & 624 \\
        CockroachDB & 6.21 & 9.44 & 6610 & 4.72 & 7.97 & 5150 \\
        Cassandra   & 1.29 & 2.70 & 142 & 1.36 & 2.99 & 124 \\
        SurrealDB   & 3.99 & 4.97 & 1311 & 2.96 & 3.78 & 1047 \\
        \midrule
        Mean & 3.67 & 6.39 & 2274 & 3.06 & 5.94 & 1870~\textcolor{green!50!black}{($\downarrow$21.6\%)}  \\
        \bottomrule
    \end{tabular}
\end{table}

\subsection{Q4: Parsing DDLs}\label{sec:ddl}

In this section, we explore whether \sqlflex{} can be extended beyond query parsing to support Data Definition Language (DDL) statements. As highlighted by the SchemaPile~\cite{till24schemapile} work, which collected database schemas, parsing DDL statements in a multi-dialect setting also remains a major challenge. Among DDL statements, \codeIn{CREATE TABLE} statements are particularly difficult to parse due to the large number of dialect-specific features, including diverse data types and DBMS-specific keywords, as observed by SchemaPile~\cite{till24schemapile}. Consequently, we focus our evaluation on the \codeIn{CREATE TABLE} DDL type. \looseness=-1

\paragraph{Extending \sqlflex{}}
Extending \sqlflex{} to support \codeIn{CREATE TABLE} statements primarily required engineering effort, while using the same hybrid segmentation algorithm. Specifically, this extension involved adding the corresponding SQL-92 grammar rules, AST node definitions, segmentation prompts, and mappings between segmented components and AST nodes. We illustrate this extension with a simplified example below, with dialect-specific features in red. Segmentation proceeds in two iterations. In the first iteration, \sqlflex{} identifies the table name (\codeIn{t}) and the column definitions, while treating the \codeIn{Engine} option as dialect-specific. In the second iteration, \sqlflex{} segments the column definition into data type (\codeIn{mediumint}), column name (\codeIn{id}), and column constraints (\codeIn{NOT NULL}).


\begin{lstlisting}[style=inline]
CREATE TABLE t (id @mediumint@ NOT NULL) @Engine=InnoDB@;
\end{lstlisting}

\paragraph{Methodology}

We followed the methodology of SchemaPile~\cite{till24schemapile} to assess the correctness of parsing results. Specifically, we used \pglast{} as the reference parser and treated its parse results as the ground truth, using the \codeIn{CREATE TABLE} statements from the SchemaPile dataset. After excluding statements that failed to be parsed by \pglast{} and applying keyword-based deduplication, 18,139 \codeIn{CREATE TABLE} statements in the PostgreSQL dialect remained. From the parsed ASTs, we extracted and compared the following elements: table names with identifier delimiters stripped, column names appearing in column definitions, and the counts of \codeIn{NOT NULL}, \codeIn{UNIQUE}, \codeIn{PRIMARY KEY}, and \codeIn{FOREIGN KEY} constraints. For each element, we calculated the percentage of ASTs whose extracted results match the ground truth. \looseness=-1


\paragraph{DDL-parsing effectiveness (Q4)}
We present the results in Table~\ref{tb:ddl}. \sqlflex{} achieves consistently high correctness across all elements, exceeding 96.41\% in every category, and outperforms \sqlglot{} on four out of the six elements, demonstrating the applicability of \sqlflex{} to parsing DDL statements. We observed that \codeIn{CREATE TABLE} statements have a generally high success rate, partly because they are less challenging than \codeIn{SELECT} statements, which tend to contain more dialect-specific features. Although \sqlflex{} performs slightly worse than \sqlglot{}$_d$ on column definitions and foreign key constraints, the difference is only at about 1\%. Moreover, when considering the percentage of queries for which all elements are parsed correctly, \sqlflex{} has a substantially higher correctness rate of 93.79\%, compared to 80.69\% for \sqlglot{}$_d$.

\begin{table}[]
    \centering
    \footnotesize
    \caption{Correctness of parsers on \codeIn{CREATE TABLE} statements. Bold indicates the best. (PK: Primary Key, FK: Foreign Key)}
        \vspace{-8pt}

    {
    \begin{tabular}{@{}lrrrrrr@{}}
    \toprule
    \textbf{Parser} & \textbf{Tables} & \textbf{Cols} & \textbf{NotNull} & \textbf{Unique} & \textbf{PK} & \textbf{FK} \\
    \midrule
    \sqlflex{}       
      & \textbf{99.75\%} 
      & 97.38\% 
      & \textbf{99.46\%} 
      & \textbf{99.72\%} 
      & \textbf{99.80\%} 
      & 96.41\% \\
    \sqlglot{}$_d$ 
      & 97.41\% 
      & \textbf{97.64\%} 
      & 88.50\% 
      & 89.03\% 
      & 97.64\% 
      & \textbf{97.64\%} \\

    \sqlglot{}$_s$ 
      & 90.86\% 
      & 91.10\% 
      & 82.29\% 
      & 83.21\% 
      & 91.10\% 
      & 91.10\% \\
    \bottomrule
    \end{tabular}
    }
    \label{tb:ddl}
\end{table}


\section{Discussion and Limitation}\label{sec:discussion}

\paragraph{Effectiveness}
To further improve the effectiveness of \sqlflex{}, several directions can be explored. First, retrieval-augmented generation may help resolve ambiguities during segmentation by retrieving relevant documentation from the target SQL dialect~\cite{zhong2025testingdatabasesystemslarge}. Second, fine-tuning the LLM could improve its hierarchical reasoning capabilities~\cite{jiang2025hibenchbenchmarkingllmscapability}. However, this approach is less scalable and presents challenges in curating representative datasets across diverse dialects~\cite{pourreza2024SQLGENBridgingDialect}. Third, although prompt engineering was not a primary focus of our work, more refined prompting strategies may further improve segmentation effectiveness. Lastly, as LLMs continue to advance, we expect the effectiveness of \sqlflex{} to grow accordingly. Given that current LLMs are autoregressive, we believe the sequential nature of our approach will remain relevant.

\paragraph{Efficiency}
While the parsing efficiency of \sqlflex{} does not yet match that of rule-based parsers, it remains suitable for non-interactive use cases or scenarios involving repeated operations on the same query, such as integrating SQL linting into CI/CD pipelines. The main efficiency bottleneck lies in the large number of LLM calls required. We expect the efficiency of \sqlflex{} to improve, as various techniques to accelerate LLM inference continue to be proposed~\cite{efficientlyScaling23, kwon2023efficient, promptcahche24}. Such advancements are also anticipated by other LLM-powered interactive systems, such as Text-to-SQL tools~\cite{tian2024SQLucid, tian-etal-2023-interactive}. Beyond model-level optimization, \sqlflex{} can further reduce the number of LLM calls by extending the base grammar with additional grammar rules for common SQL features, thus avoiding redundant segmentations. For instance, when inspecting queries that required LLM-based segmentation, we found that some common, but non-standard features like the \codeIn{::} type-casting operator and \codeIn{LIMIT} clause were often contained. By adding such features to the grammar, the performance could be further optimized in practice. In an additional experiment, where we added the \codeIn{::} and \codeIn{LIMIT} features to the SQL:2003 grammar, we could further reduce the number of LLM calls by 36.33\% on the PostgreSQL dataset, for example. Moreover, existing grammar inference techniques that automatically derive context-free grammars from parse trees~\cite{gopinath2020Mining, schroeder2022mining} could enable \sqlflex{} to cache recurring query patterns or dynamically integrate newly learned rules into its grammar-based parser, further minimizing the need for LLM invocations.

\paragraph{Soundness}

Unlike traditional parsing algorithms that provide formal correctness guarantees, \sqlflex{} adopts a best-effort hybrid parsing approach due to its usage of LLMs. However, achieving full semantic correctness is inherently difficult in a multi-dialect setting without explicit dialect knowledge, and this challenge also affects traditional multi-dialect parsers such as \sqlglot{}. A representative example is the \codeIn{||} operator: in MySQL it denotes logical OR, while in PostgreSQL and SQLite it performs string concatenation. Moreover, PostgreSQL assigns \codeIn{||} lower precedence than arithmetic operators, whereas SQLite assigns it higher precedence. As a result, the expression \codeIn{3 * 2 || `7'} evaluates differently across the three DBMSs (\codeIn{true}, 81, and 67 for MySQL, SQLite, and PostgreSQL, respectively). Notably, SQLGlot assigns \codeIn{||} the lowest precedence uniformly across dialects, which is incorrect for SQLite. As shown in our evaluation, traditional parsers such as \sqlglot{} also suffer from parsing and correctness issues due to the diversity  and evolution of SQL dialects. In contrast, \sqlflex{} achieves high correctness on the PostgreSQL dataset and proves effective in practical applications such as SQL linting and test-case reduction, where constructing a structurally sound AST across diverse dialects is more important than strict semantic guarantees. Finally, \sqlflex{}’s validation and repair mechanisms further improve its robustness in practice.

\paragraph{Use cases}
{While this paper reports the results of applying \sqlflex{} to only two applications, \sqlflex{} is, in principle, applicable to any AST-based SQL analysis or rewriting workflow rather than being limited to task-specific scenarios. Additional applicable use cases include SQL-level optimization (\eg{} Calcite faces similar
dialect-related parsing challenges~\cite{calcite-doris-issue}), data lineage and provenance analysis~\cite{miao19explaining, miao20irex}, and evaluations of Text-to-SQL systems~\cite{li2024DawnNaturalLanguage, zhong-etal-2020-semantic}.
Some applications, such as SQL-level optimization, are correctness-critical. While SQLFlex lacks formal correctness guarantees, it can be paired with SQL solvers~\cite{verieql24he} to strengthen semantic assurances, as demonstrated by recent agentic query optimization frameworks that use LLMs to generate optimized queries~\cite{song2025quitequeryrewriterules}. Since SQL solvers also face challenges in multi-dialect support, further research on making them dialect-aware might be needed.
The strong practical demand for cross-dialect parsing is further
reflected in the significant engineering effort behind open-source
projects, such as SQLGlot, SQLFluff, and Calcite, whose issue trackers~\cite{calcite-doris-issue, sqlfluff-h2-issue, sqlglot-dremio-issue} indicate that dialect coverage remains a persistent pain
point. Yet, despite this clear demand, prior research has offered
few broadly applicable solutions. SQLFlex aims to fill this gap by
providing a dialect-agnostic parsing framework that benefits SQL
tooling beyond the use cases presented.}

\section{Related Work}

\paragraph{SQL parsers}
Existing SQL parsers can be categorized into generated parsers and hand-written parsers. Generated parsers are automatically built from formal grammar specifications. For example, ANTLR~\cite{antlr} generates parsers that use the LL(*) algorithm~\cite{llstar2011Parr}. Some DBMSs, such as PostgreSQL, use YACC-style toolkits~\cite{yacc} for their built-in parsers. Mühleisen et al. proposed using Parsing Expression Grammars~\cite{ford2004PEG} as a more extensible and efficient parser generator for DuckDB~\cite{muhleisen2025runtime}. Hand-written parsers offer greater flexibility for dialect-specific syntax. A state-of-the-art example is \sqlglot{}~\cite{SQLGlot}, a widely used multi-dialect SQL parser with over 8,000 GitHub stars, supporting more than 30 dialects. However, such tools require substantial manual effort to maintain and extend. \codeIn{sqlparse}~\cite{sqlparse} is a non-validating parser that builds approximate parse trees, but is unreliable for query rewriting~\cite{dintyalaSQLCheckAutomatedDetection2020}. To the best of our knowledge, \sqlflex{} is the first work that integrates LLM with grammar-based parsing to achieve dialect-agnostic query parsing. \looseness=-1


\paragraph{SQL dialects}
Multiple approaches have addressed the SQL dialect problem in various applications. In DBMS testing, tools like Sedar~\cite{fuSedarObtainingHighQuality2024} and QTRAN~\cite{lin2025qtran} leverage LLMs to translate test cases across different dialects.
Recent Text-to-SQL benchmarks like MiniDev~\cite{liCanLLMAlready} (extending to MySQL and PostgreSQL) and Spider 2.0~\cite{lei2024spider} (including BigQuery and Snowflake) now incorporate more dialects in addition to SQLite. Closely related are tools like SQL-GEN~\cite{pourreza2024SQLGENBridgingDialect}, which synthesizes dialect-specific training data, and new architectures like MOMQ's Mixture-of-Experts model~\cite{lin2024momqmixtureofexpertsenhancesmultidialect} and Exec-SQL's use of execution feedback~\cite{zhang2025exesqlselftaughttexttosqlmodels}, all aimed at improving multi-dialect capabilities of LLMs. Dialect translation is crucial for applications like data migration. CrackSQL~\cite{zhou2025cracking} is a state-of-the-art tool combining rule-based and LLM-based approaches for translation. In contrast to these works that focus on generating or translating queries between dialects, \sqlflex{} addresses the foundational challenge of dialect-agnostic SQL parsing.

\paragraph{Query rewriting}
Many applications can be categorized as query rewriting under our formulation. Beyond the discussed use cases, widely researched rewriting tasks include query optimization and DBMS testing. In query optimization, predefined rules are applied to improve query performance~\cite{graefe1987exodus}. More recent approaches, such as WeTune~\cite{wangWeTuneAutomaticDiscovery2022} and LearnedRewrite~\cite{zhouLearnedQueryRewrite2021}, automatically discover and apply rewrite rules. LLM-R$^2$~\cite{li2024LLMR2LargeLanguage} extends this by using LLMs to explore more effective rewrite rules. QueryBooster~\cite{baiQueryBoosterImprovingSQL2023} introduces a domain-specific language for user-defined optimization rules. In DBMS testing, numerous works rewrite queries to detect bugs. Mutation-based fuzzers like WingFuzz~\cite{jie2024WingFuzz} and LEGO~\cite{liang2023SequenceOriented} mutate queries to generate test inputs. Some approaches, like EET~\cite{jiang24eet}, AMOEBA~\cite{liuAutomaticDetectionPerformance2022}, and SQLancer~\cite{Rigger2020TLP, Rigger2020NoREC}, generate equivalent queries to check for consistent results. Commonly, these approaches rely on parsers to analyze and modify query ASTs. \sqlflex{} addresses a key limitation by enabling dialect-agnostic query parsing, broadening the applicability of query rewriting tools.


\section{Conclusion}
Existing SQL parsers often require significant manual effort to support diverse SQL dialects, limiting the applicability of query analysis and rewriting tools. In this paper, we have presented \sqlflex{}, a dialect-agnostic query rewriting framework. Our key idea is to integrate grammar-based parsing with LLM-based segmentation for dialect-agnostic query parsing. We introduced clause-level and expression-level segmentation to decompose the hierarchical structure of queries into sequential tasks. We also proposed validation methods to improve reliability. Our evaluation has shown that \sqlflex{} is practical in real-world SQL linting and test-case reduction tasks. Additionally, \sqlflex{} can parse 91.55\% to 100\% of the queries across eight dialects, outperforming baselines. We believe \sqlflex{} can be applied to a broad range of applications, and requires minimal manual adaptation for dialect-specific features.

\begin{acks}
We would like to thank the anonymous reviewers for their constructive comments and suggestions. This project is supported by the Ministry of Education, Singapore, under the Academic Research Fund Tier 1 (FY2023).
\end{acks}

\clearpage
\bibliographystyle{ACM-Reference-Format}
\bibliography{main}

\appendix
\section{Additional Approach Details}
\subsection{Grammar Alternatives and Reptitions}
Some grammar rules specify repetitions, where an element is allowed to appear zero or multiple times. In ANTLR, we use the symbol \codeIn{*} to denote such repetitions. In the listing below, the \codeIn{projections} rule uses the pattern \codeIn{("," projection)}, allowing multiple \codeIn{projection} symbols separated by commas. Some grammar rules define alternatives, offering a choice between multiple possible symbols. In ANTLR, alternatives are specified using the \codeIn{``|''} symbol. For instance, the \codeIn{projection} rule provides two alternatives: \codeIn{column} and \codeIn{expr}, meaning a \codeIn{projection} can be either a simple column name or a more complex expression.
\begin{lstlisting}[style=inline]
projections: projection ("," projection)*;
projection: column | expr;
\end{lstlisting}

In clause-level segmentation, we handle both repetitions and alternatives as special cases. For repetitions, the LLM outputs the full sequence, which we then split using delimiters. For example, in a \codeIn{SELECT} clause like \codeIn{``SELECT col, 1 !> 2''}, the LLM may output the segment \codeIn{``col, 1 !> 2''}, corresponding to the \codeIn{projections} rule. We then split this by commas and enqueue each element (\codeIn{``col''} and \codeIn{``1 !> 2''}) under the \codeIn{projection} rule for further processing. For alternatives, the LLM selects one of the possible forms. For example, when segmenting \codeIn{``1 !> 2''} under the \codeIn{projection} rule, the LLM identifies it as an expression rather than a column name. As a result, we enqueue \codeIn{``1 !> 2''} with the \codeIn{expr} rule, which \sqlflex{} proceeds with expression-level segmentation in the next step. Additionally, the \emph{mutually exclusive validation} also applies to such scenarios where alternatives are defined (\eg{} segmentation output for \codeIn{projection} cannot be interpreted as both a column name and an expression).

\subsection{Conservative Anchor Matching}
To identify anchors reliably, we adopt a conservative matching strategy to minimise incorrect matches that could lead to parsing errors. Specifically, we only match anchors that are space-separated. For example, in \codeIn{``year < 2025''}, the operator \codeIn{``<''} is surrounded by spaces and thus qualifies as a match. In contrast, we skip matching anchors in expressions written as \codeIn{``year<2025''}, as such compact forms may reflect dialect-specific features (\eg{} the operator \codeIn{``!<''} includes \codeIn{``<''}, but since \codeIn{``<''} lacks surrounding spaces, matching is avoided). Additionally, we handle phrase-like operators such as \codeIn{``IS NOT''} through exact matching. As a result of this conservative approach, some known operators may remain unmatched and appear within mask tokens, though these expressions might still be parseable in subsequent steps.

\subsection{Context-sensitive Anchors}
Certain anchors used in expression-level segmentation are context-sensitive, that is, they do not function as universal operators like \codeIn{``AND''}, but act as operators only in specific contexts. For example, consider the \codeIn{CAST} function: in an expression like \codeIn{``CAST(1.0 AS INT)''}, the keyword \codeIn{``AS''} serves as a type-cast operator. However, outside of such function calls, \codeIn{AS} typically denotes an alias (\eg{} \codeIn{``SELECT col AS alias''}). Due to this ambiguity, we treat \codeIn{AS} as an anchor only when the current expression node has a parent that is the \codeIn{``CAST''} function. In such cases, the context makes it clear that \codeIn{AS} is being used as a type-cast operator rather than an alias specifier. While such context-sensitive anchors are rare in the SQL-92 grammar, we believe they are more prevalent in more comprehensive grammars, such as specifying the anchors in window functions.

\section{Additional Evaluation Details}
\subsection{Dataset construction}
\paragraph{Standalone evaluation}
To build the dataset for the standalone evaluation, we extracted \codeIn{SELECT} statements from the test files. For DuckDB, PostgreSQL, and CockroachDB, we used SQUaLity~\cite{zhong2024UnderstandingReusingTest}, a DBMS test suite analyzer to parse test files from the two systems. For other DBMSs not supported by SQUaLity, we matched for any string starting with \codeIn{"SELECT"}. To ensure syntactic validity, we excluded queries that failed to parse using dialect-specific parsers. For H2, since we were unable to obtain a ready-to-use H2-specific parser, we used JSQLParser, which supports H2 syntax. Since many of the extracted queries had similar structures or features, we applied a keyword-based deduplication heuristic to retain a diverse subset. SQL parsing depends primarily on keywords~\cite{sqlHasProblems24Shute}, making keyword diversity a useful proxy for syntactic variety. Specifically, for each DBMS, we compiled a list of SQL keywords and used regular expressions to extract the keyword set from each query. Queries with identical keyword sets were treated as duplicates, and we retained only the longest one, assuming it to be the most complex. For SurrealDB, as we were unable to obtain its keyword list, we kept all queries in its dataset.

\subsection{AST-level Parsing Effectiveness}
We introduce \emph{AST Round-trip (AST-RT)}, an AST-level round-trip property to gain additional insights on the parsing effectiveness of \sqlflex{} against the baselines. Similarly, AST-RT is defined based on the following parsing and printing steps:
\begin{align}
    \text{AST}_0 &= \text{Parse}(Q_0),\quad Q_1 = \text{Print}(\text{AST}_1) \\
    \text{AST}_1 &= \text{Parse}(Q_1),\quad Q_2 = \text{Print}(\text{AST}_2)
\end{align}
\emph{AST Round-trip (AST-RT)} measures AST-level consistency by checking that the AST remains unchanged after two parsing operations (\ie{} $\text{AST}_0 = \text{AST}_1$). AST-RT demonstrates whether \sqlflex{} consistently generates structurally stable ASTs, which is useful for users implementing rewrite functions. AST-RT avoids disadvantaging \pglast{} and \sqlglot{} due to certain normalizations. For example, \pglast{} combines type-casting operations into a unified node type, and direct AST comparisons avoid false alarms.

Table~\ref{tb:ast-rt} shows the results in AST-RT rates, with the geometric mean reported in the last row. The standard deviations of \sqlflex{} in AST-RT rates are slightly higher than Q-RT rates, but remain below 0.7\%. \sqlflex{} overall outperforms all baselines across most datasets, with only a marginal 3\% gap behind \sqlglot{}$_s$ on the DuckDB dataset. In general, AST-RT scores tend to be slightly lower than Q-RT scores. For \sqlglot{} and \pglast{}, this is often caused by normalization during pretty-printing. For example, in \sqlglot{}, an \codeIn{IF} node in the original AST is converted into a \codeIn{CASE} expression in the printed query, which then leads to a different node structure when the query is re-parsed. This lower AST-RT is also expected for the LLM-based parser and \sqlflex{}. Since LLMs are inherently probabilistic, they may produce slightly different ASTs even for the same input, especially when the query is complex. For \sqlflex{}, such discrepancies mainly occur in complex expressions without anchors, such as window functions. Window functions are not supported in SQL-92, but are common in DuckDB, PostgreSQL, and CockroachDB datasets. Without anchors, the segmenter needs to process the expression recursively, and small differences in interpretation can lead to AST mismatches. Nonetheless, the printed queries remain largely consistent, as reflected in the high Q-RT scores, indicating that re-executability is preserved. 


\begin{table}[]
    \centering
    \setstretch{1.1}
    \footnotesize
    \caption{AST Round-Trip Rates. Bold indicates the best score for each dataset (excluding \pglast{} on PostgreSQL)}
    \begin{tabular}{@{}lccccc@{}}
    \toprule
    \textbf{Dataset} & \textbf{LLM} & \textbf{\pglast{}} & \textbf{\sqlglot{}$_s$} & \textbf{\sqlglot{}$_d$} & \textbf{\sqlflex{}} \\
    \midrule
         DuckDB & 26.00\% & 72.13\% & \textbf{92.16}\% & 92.11\% & 89.24\% \\
         PostgreSQL & 25.35\% & 100.00\% & 84.92\% & 86.74\% &  \textbf{90.25\%} \\
         MySQL & 33.69\% & 77.70\% & 81.42\% & 93.36\% & \textbf{97.24\%} \\
         ClickHouse & 24.83\% & 46.96\% & 64.27\% & 88.93\% & \textbf{93.66\%} \\
         H2 & 26.26\% & 81.01\% & 93.30\% & - & \textbf{95.07\%} \\
         CockroachDB & 32.21\% & 69.54\% & 66.84\% & - & \textbf{88.34\%} \\
         Cassandra & 38.00\% & 74.00\% & 76.00\% & - & \textbf{99.56\%} \\
         SurrealDB & 25.64\% & 32.76\% & 38.46\% & - & \textbf{92.40\%} \\
     \midrule
        Mean & 28.48\% &  65.84\% & 72.30\% & 90.25\% & \textbf{93.15\%} \\
    \bottomrule
    \end{tabular}
    \label{tb:ast-rt}
\end{table}

\section{Use Case: Mutation Testing}
We present an additional use case of \sqlflex{}, mutation testing, beyond the two use cases we have presented in the main paper.

Mutation testing is a well-established approach used to assess the effectiveness of test suites. It involves generating multiple versions of a program, known as mutants, where each mutant contains a small, intentional fault introduced by a simple syntactic change in the original code. The quality of a test suite is then evaluated based on its ability to detect and distinguish these mutants from the original program~\cite{budd1980mutation, papadakis2019mutation, original_mutation_testing}.

In the context of SQL, mutation testing is a use case of query rewriting. A SQL test suite typically consists of the database schema, the data, a set of \codeIn{SELECT} queries, and their expected results. Such test suites are essential in various use cases, including validating the reliability and correctness of database applications, SQL query engines, solutions to online SQL coding platforms, and evaluation benchmarks. The queries in the test suite can be mutated (\ie{} rewritten) prior to execution to assess the adequacy of the test suite. A SQL mutant is said to be \emph{dead} when the output produced by its execution differs from the output produced by the original \codeIn{SELECT} query. This indicates that the test data is capable of distinguishing between the original and mutated queries, while any surviving mutant indicates a potential gap or weakness in the test data.

Although mutation testing for SQL queries was introduced as early as 2006~\cite{tuya2006SQLMutationToolGenerate}, it has not seen widespread adoption. A key barrier is the lack of dialect-agnostic query support across different database engines, making it difficult to build generalised testing tools. Most existing approaches perform mutations at the raw query string level, which limits structural awareness and leads to fragile or semantically invalid transformations. In contrast, SQLFlex enables mutation testing at the AST level by generating unified ASTs from SQL queries. This allows structure-aware and semantically valid query mutations to be applied. For example, SQLFlex can apply a mutation rule that modifies the type of a join clause (if present in the query), such as changing \codeIn{``LEFT JOIN''} to \codeIn{``RIGHT JOIN''}.


\subsection{Baselines}
SQLMutation \cite{SQLMuation} is a prominent tool designed to generate SQL mutants by applying mutations to the \texttt{SELECT} queries within a test suite. However, as SQLMutation is not available as a standalone tool and offers limited control over fine-grained mutant generation, we implemented a similar mutation tool. Following the standalone evaluation baseline, we implemented a baseline mutation testing tool based on the PostgreSQL-based parser, \codeIn{pglast}. This gave us greater flexibility in selecting specific mutants for a more meaningful and controlled comparison with our SQLFlex-based mutation testing tool. 


\subsection{Implementation}

We systematically selected twelve mutation rules (\ie{} query rewriting rules) from the rules defined by Tuya et al.~\cite{tuya2007mutating}. Specifically, we excluded mutation rules that require schema-level information, as our evaluation focuses exclusively on mutations that can be derived directly from the \codeIn{SELECT} statements themselves. We also excluded mutation rules on subqueries to reduce the complexity of the tool for the purpose of evaluation. We implemented the twelve rules for the PostgreSQL-based mutation tool and the SQLFlex-based mutation tool.

\subsection{Dataset}
SQL mutation testing is particularly relevant in contexts such as online platforms that host SQL coding exercises, Text-to-SQL benchmarks, and DBMS test suites, where ensuring the semantic uniqueness of query outputs is critical. We selected four Text-to-SQL benchmarks to evaluate the mutation testing tools due to their open-source availability and close resemblance to real-world queries and data. Specifically, the selected benchmarks are BIRD~\cite{li2024can}, MiniDev~\cite{li2024can}, Spider 1.0~\cite{spider1}, and Spider 2.0~\cite{lei2024spider}.

\subsection{Results}

The presence of surviving mutants in a test suite highlights its inadequacy, where a higher number of surviving mutants generally indicates greater deficiencies in the test data. These mutants can often be eliminated by adding more targeted test cases to the suite. However, a larger number of surviving mutants does not necessarily mean that a mutation testing tool is more effective. Ideally, we should also consider the significance of each surviving mutant. For example, whether eliminating one mutant with additional test data also helps eliminate others. Nevertheless, a larger number of surviving mutants can still be a useful indicator for revealing differences between mutation testing tools.


\begin{table}[]
    \centering
    \footnotesize
    \caption{Mutation testing results}
    \begin{tabular}{@{}lrlr@{}}
        \toprule
        \textbf{Dataset} & \textbf{\#Queries} & \textbf{Approach} & \textbf{\#Living Mutants} \\
        \midrule
        \multirow{2}{*}{BIRD (SQLite)} & \multirow{2}{*}{766} & SQLFlex & \textbf{3,551} \\
        & & \pglast{} & 3,413 \\
        \midrule
        \multirow{2}{*}{Spider 1.0 (SQLite)} & \multirow{2}{*}{890} & SQLFlex & \textbf{4,464} \\
        & & \pglast{} & 4,127 \\
        \midrule
        \multirow{2}{*}{Spider 2.0 (Snowflake)} & \multirow{2}{*}{121} & SQLFlex & \textbf{234} \\
        & & \pglast{} & - \\
        \midrule
        \multirow{2}{*}{MiniDev (PostgreSQL)} & \multirow{2}{*}{500} & SQLFlex & 2,419 \\
        & & \pglast{} & \textbf{2,426} \\
        \midrule
        \multirow{2}{*}{MiniDev (MySQL)} & \multirow{2}{*}{500} & SQLFlex & \textbf{1,958} \\
        & & \pglast{} & 24 \\
        \midrule
        \multirow{2}{*}{MiniDev (SQLite)} & \multirow{2}{*}{500} & SQLFlex & \textbf{2,377} \\
        & & \pglast{} & 2,327 \\
        \bottomrule
    \end{tabular}
    \label{tab:mutation-comparison}
\end{table}

As shown in Table~\ref{tab:mutation-comparison}, the mutation tool based on SQLFlex generally produces more surviving mutants, with the exception of the MiniDev dataset in the PostgreSQL dialect. This exception is expected, as the PostgreSQL-specific parser is inherently better suited for this dataset. Notably, on the Spider 2.0 dataset with the Snowflake engine, SQLFlex demonstrates a clear advantage. Specifically, it successfully generates executable mutant queries, whereas the PostgreSQL-based tool introduces errors by converting backticks to double quotes during pretty-printing. This normalization makes the queries invalid on Snowflake, resulting in no surviving mutants.

\received{October 2025}
\received[revised]{January 2026}
\received[accepted]{February 2026}
\end{document}